\documentclass[12pt,preprint]{aastex}

\pdfoutput=1

\slugcomment{\today}

\shorttitle{HST view of YSOs in the LMC} \shortauthors{Vaidya et al.}

\begin{document}

\title{An {\it HST} View of the Interstellar Environments of Young Stellar Objects in the Large Magellanic Cloud}


\author{
        Kaushar Vaidya\altaffilmark{}, 
        You-Hua Chu\altaffilmark{},
	Robert A. Gruendl\altaffilmark{}}
\affil{Department of Astronomy, University of Illinois at Urbana-Champaign, 1002 West Green Street, Urbana, IL 61801, USA}

\author{
       C.-H. Rosie Chen\altaffilmark{}}
\affil{Department of Astronomy, University of Virginia, 530 McCormick Road, Charlottesville, VA 22904, USA}

\and
\author{ 
        Leslie W. Looney\altaffilmark{}}
\affil{Department of Astronomy, University of Illinois at Urbana-Champaign, 1002 West Green Street, Urbana, IL 61801, USA}



\begin{abstract}
We have used archival {\it HST} H$\alpha$ images to study the immediate environments 
of massive and intermediate-mass young stellar object (YSO) candidates in the Large 
Magellanic Cloud (LMC).  The sample of YSO candidates, taken from \citet{gruendl08}, 
was selected based on {\it {\it Spitzer}} IRAC and MIPS observations of the entire 
LMC and complementary ground-based optical and near-infrared observations.  
We found {\it HST} H$\alpha$ images for 99 YSO candidates in the LMC, of which 82 
appear to be genuine YSOs.  More than 95\% of the YSOs are found to be associated with 
molecular clouds.  YSOs are seen in three different kinds of environments in the 
H$\alpha$ images: in dark clouds, inside or on the tip of bright-rimmed dust pillars, 
and in small \ion{H}{2} regions.  Comparisons of spectral energy distributions for
YSOs in these three different kinds of environments suggest that YSOs in dark clouds
are the youngest, YSOs with small \ion{H}{2} regions are the most evolved, and YSOs 
in bright-rimmed dust pillars span a range of intermediate evolutionary stages.  
This rough evolutionary sequence is substantiated by the presence of silicate 
absorption features in the {\it Spitzer} IRS spectra of some YSOs in dark clouds 
and in bright-rimmed dust pillars, but not those of YSOs in small \ion{H}{2} regions.  
We present a discussion on triggered star formation for YSOs in bright-rimmed 
dust pillars or in dark clouds adjacent to \ion{H}{2} regions.
As many as 50\% of the YSOs are resolved into multiple sources in high-resolution
{\it HST} images.  This illustrates the importance of using high-resolution images to
probe the true nature and physical properties of YSOs in the LMC.

\end{abstract}




\section{Introduction}

The {\it Spitzer Space Telescope}, with its high angular resolution and sensitivity 
at mid-infrared wavelengths, has made it possible for the first time to survey 
massive young stellar objects (YSOs) in nearby external galaxies.  In particular, 
in the Large Magellanic Cloud (LMC), because of its close proximity (50 kpc; 
\citealt{feast99}) and low inclination ($\sim$ 30$\arcdeg$; \citealt{nikolaev04}), 
massive and intermediate-mass YSOs can be resolved by {\it Spitzer} and inventoried 
throughout the entire galaxy.  The LMC was observed by {\it Spitzer} under a Legacy 
Program, Surveying the Agents of Galaxy Evolution (SAGE), that mapped the central 
7\arcdeg\ $\times$ 7\arcdeg\ area of the galaxy \citep{meixner06}. \citet{gruendl08} 
made use of the archival {\it Spitzer} data from the SAGE survey along with 
complementary ground-based optical and near-infrared observations and identified a 
comprehensive sample of massive and intermediate-mass YSO candidates of the entire LMC.
This sample consists of a total 855 {\it definite}, 317 {\it probable}, and 213 
{\it possible} massive ($>$ 10 $M_\sun$) and intermediate-mass ($>$ 4 $M_\sun$) 
YSO candidates.  Most of these YSO candidates are found to be concentrated in or 
around molecular clouds or \ion{H}{2} regions \citep{gruendl08}.

While {\it Spitzer} enables the detection of individual massive and intermediate-mass 
YSOs in the LMC \citep[e.g.,][]{chu05,jones05,caulet08,whitney08,gruendl08}, the 
angular resolution of {\it Spitzer} is not sufficient to resolve multiple sources 
within 1--2$\arcsec$ or allow a close look at the environments of these massive YSOs.  
The {\it Hubble Space Telescope (HST)} on the other hand, with an angular resolution 
ten times higher than that of {\it Spitzer}, reveals sub-arcsec size features and 
hence is very useful for a detailed examination of the environments of these massive YSOs.  

We have therefore searched through the {\it HST} archive for continuum 
and H$\alpha$ images that cover LMC YSO candidates from the above catalog. 
In many cases, the exposure times of the continuum images are short and hence 
not very useful, as noted by \citet{chen08}.  The H$\alpha$ images have longer 
exposure times and are very useful in revealing interstellar environments of
YSOs in \ion{H}{2} complexes where the interstellar gas is photoionized
by antecedent massive stars.  For example, ionization fronts on the surface
of dense molecular clouds can be recognized by the sharply enhanced H$\alpha$ 
surface brightness.   H$\alpha$ images are also useful in revealing 
circumstellar environments of massive YSOs.  For example, the UV flux of a 
YSO may photoionize its circumstellar medium and form a compact \ion{H}{2}
region, or outflows from a YSO may produce observable features like Herbig-Haro 
objects \citep{heydari99,chu05}.  Thus, in this paper we make use of primarily 
the {\it HST} H$\alpha$ images to examine the immediate surroundings of the LMC 
YSO candidates.  The continuum images are only used to supplement the analyses of 
stellar properties for a few YSOs.

The remaining paper is organized as follows:  Section 2 describes the datasets used 
in this work and our method of analysis, Section 3 describes the environments of YSO 
candidates, Section 4 discusses the properties of small \ion{H}{2} regions discovered 
around some massive YSO candidates, Section 5 describes the mid-infrared spectral 
characteristics of some YSO candidates, Section 6 discusses the spectral energy 
distributions (SEDs) of the YSO candidates, Section 7 presents a discussion on 
triggered star formation, and finally Section 8 summarizes the paper.

\section{Datasets and Method of Analysis}

We searched the {\it HST} archive for H$\alpha$ images observed with the Wide 
Field Planetary Camera 2 (WFPC2) and the Advanced Camera for Surveys (ACS) for
a field of $\sim$ 78 degree$^2$ centered on the LMC coordinates
($\alpha$ = 05$^{h}$ 18$^{m}$, $\delta$ = $-$68$\arcdeg$ 34$\arcmin$), 
available as of 2008 August.   WFPC2 consists of four cameras, PC1, WF2, WF3, 
and WF4, among which PC1 has a field-of-view of 36$\arcsec$ $\times$ 36$\arcsec$ 
with a scale of 0\farcs045 pixel$^{-1}$, and the remaining three cameras each
have a field-of-view of 80$\arcsec$ $\times$ 80$\arcsec$ with a scale of 
0\farcs1 pixel$^{-1}$ \citep{mcmaster08}. The ACS observations were made 
in the Wide Field Channel (WFC), which has a field-of-view of 
202$\arcsec$ $\times$ 202$\arcsec$ with a scale of 0\farcs05 pixel$^{-1}$ 
\citep{boffi07}.   The H$\alpha$ filters F656N ($\lambda_c$ = 6564\AA, 
$\Delta \lambda$ = 22\AA) and F658N ($\lambda_c$ = 6584\AA, $\Delta \lambda$ = 72\AA) 
were used with WFPC2 and ACS observations, respectively.

As noted in \citet{gruendl08}, we have been imaging YSO candidates in selected 
regions throughout the LMC in the near-infrared $J$ and $K_s$ bands with the IR 
Side Port Imager (ISPI) on the Blanco 4 m telescope at the Cerro Tololo 
Inter-American Observatory.   The ISPI camera has a $10\farcm25 \times 10\farcm25$ 
field-of-view imaged with a $2048\times2048$ HgCdTe Array with 0\farcs3 pixels.  
In total, we have imaged 113 fields in four different runs, 2005 November, 
2006 November, 2007 February, and 2008 January.
During our last run in 2008 January, we specifically imaged 29 fields that 
encompassed the {\it HST} archival data used in our study of YSO candidates in 
the LMC.  These observations give us additional confirmation on the nature of the 
YSO candidates, bridging the gap between optical and mid-infrared wavelengths.  
Each field was imaged with a sequence of exposures with a telescope offset of 
$\sim$ 1$\arcmin$ between frames to aid in removal of bad pixels and to facilitate
sky subtraction and flat fielding.  For the $J$-band observations, thirteen 30 s
exposures were obtained, while at $K_s$ band, twenty-three 30 s frames (each 
consisting of two coadded 15 s exposures) were obtained.  The observations were 
non-linearity corrected, sky subtracted, and flat fielded using standard routines 
within the CIIRED and SQIID packages in IRAF.  The astrometry was performed using 
the WCSTOOLS task IMWCS and the Two Micron All Sky Survey Point Source Catalog
(2MASS PSC; \citealt{skrutskie06}).  In these ISPI images, we find that point 
sources with $m_J$ $\lesssim$ 18.5 and $m_K$ $\lesssim$ 17.6 mag are generally
detected with better than 10-$\sigma$ significance.

Following the method of \citet{gruendl08} for assessing the nature of YSO 
candidates, we simultaneously displayed ``postage stamp'' images of each 
YSO candidate (field-of-view 5$\arcmin$ $\times$ 5$\arcmin$) in the following 
wavelengths: Digitized Sky Survey red continuum, {\it HST} H$\alpha$, ISPI $J$ 
and $K_s$ bands, {\it {\it {\it Spitzer}}} IRAC 3.6, 4.5, 5.8, and 8.0 $\mu$m, 
and MIPS 24 and 70 $\mu$m, along with the SED of the YSO candidate. 
To examine the large-scale environments, we also used the H$\alpha$ images 
from the Magellanic Cloud Emission Line Survey (MCELS; \citealt{smith99}).
By examining the source morphology and environment of each YSO candidate 
in the multiwavelength images, in conjunction with its SED profile, we conclude 
that 82 of the 99 YSO candidates are most likely genuine  YSOs.  Our evaluation 
of the YSO nature of the sources is consistent with the previous assessment of 
\citet{gruendl08}, as they categorized the 82 confirmed YSO candidates as {\it definite}
YSOs and the remaining 17 YSO candidates as {\it probable} or {\it possible} YSOs, 
where {\it probable} YSOs are likely to be YSOs but also show some characteristic 
of a possible alternative nature, and {\it possible} YSOs are more likely to be 
non-YSOs but cannot be ruled out as YSOs.  
In our analysis, the majority of these 17 non-YSOs appear to be either peaks of 
diffuse emission or background galaxies (see the table of non-YSOs in the appendix).  
In the following discussion, we include only these 82 YSO candidates and refer to 
them as YSOs.   

\section{Environments of YSOs}

The {\it HST} H$\alpha$ images provide an unprecedented detailed view of 
the immediate environments of YSOs in the LMC.
It is remarkable that all 82 YSOs are found in only three kinds
of environments: (1) in a dark cloud, (2) inside or on the tip of a bright-rimmed
dust pillar, and (3) in a small \ion{H}{2} region.  Eight YSOs are found
in more than one kind of environment, e.g., a small \ion{H}{2} region
located inside a bright-rimmed dust pillar. All YSOs are illustrated by
20$''\times 20''$ (5.0 pc $\times$ 5.0 pc) images presented in Figure~1. 
We also show the SEDs of YSOs in Fig.~1.   
For YSOs with neighboring sources within 1\arcsec, the
identification of their optical counterparts in the {\it
HST} images is not straightforward because of the combined
uncertainties in astrometry, and it is necessary to
bootstrap through images at intermediate wavelengths.
Therefore, in Fig.~1
we have included ISPI $J$ and $K_s$, and IRAC 3.6 $\mu$m
images, in addition to the {\it HST} H$\alpha$ images. 
The IRAC 3.6 $\mu$m images show the YSOs
as well as a small number of background stars. 
The alignment between the IRAC and ISPI images can be fine-tuned with these 
background stars, and a YSO's near-infrared counterpart can be identified 
by its being brighter in $K_s$ than $J$ compared with normal stars.  
The larger number of background stars in $J$ makes it easier to align 
the ISPI images with the {\it HST} H$\alpha$ image and identify a YSO's 
optical counterpart.  
General remarks of these three categories are given below.

\noindent
{\bf Dark Clouds} -- 
49 YSOs are found to be associated with dark clouds.
Among these, 34 show no detectable optical counterparts and
15 show faint optical counterparts in {\it HST} H$\alpha$ images;
their images presented in Fig.~1 are labeled as categories 1a, and 1b, respectively.
It can be seen that these YSOs are in highly dusty environments,
with dust features extending over several parsecs and 
sometimes tens of parsecs.  Some YSOs are located near ionization 
fronts, as indicated by bright H$\alpha$ emission on the surface
of dark clouds.  Note that diffuse ionized gas sometimes exists 
along the line of sight, and in these cases dark clouds can be diagnosed
only in images larger than those shown in Fig.~1, viewed with 
adjustable contrast. Two examples are the  
YSO J051351.51-672721.9 and YSO J053838.45-690418.3.

\noindent
{\bf Bright-Rimmed Dust Pillars} --
19 YSOs are associated with bright-rimmed dust pillars, of which eight are 
also associated with either dark clouds or small \ion{H}{2} regions.
These YSOs are labeled as category 2 in Fig.~1.  A great majority of these YSOs are 
projected near the tips of bright rims of dust pillars; only three YSOs 
are projected inside dust pillars.  The bright-rimmed dust pillars show 
different morphologies and sizes, but all of them are pointing toward
nearby OB associations, the source of ionizing radiation.  Three of the 
YSOs associated with dust pillars are also found to be in compact 
\ion{H}{2} regions.

\noindent
{\bf \ion{H}{2} Regions} --
Among the 22 YSOs in this category, the existence of small \ion{H}{2} regions 
ranges from ``obviously seen'' to ``implied by generalization,''  as discussed
in detail later in Section 4.  
Briefly, eight YSOs are surrounded by resolved \ion{H}{2}
regions of sizes up to 7\farcs5 ($\sim$ 1.8 pc); four YSOs appear more 
extended than the point spread function (PSF) in the {\it HST} H$\alpha$ 
images, suggesting the existence of barely resolved \ion{H}{2} regions; 
four YSOs have observed fluxes in the H$\alpha$ band higher than the 
expected stellar continuum fluxes, indicating excess H$\alpha$ emission 
from ionized circumstellar gas; the remaining six YSOs either do not 
show significant excess H$\alpha$ emission or have no optical photometric data to 
assess the expected continuum fluxes.  The images of YSOs with 
well-resolved, barely resolved, and unresolved \ion{H}{2} regions shown in 
Fig.~1, are labeled as categories 3a, 3b, and 3c, respectively.

To examine the molecular environment of these YSOs, we have compared 
the locations of YSOs with the NANTEN CO (J=1--0) survey of the LMC,
made with a 4-m telescope for a beamsize of 2\farcm6 \citep{fukui01,fukui08}.
We find that 70 YSOs are superposed on giant molecular clouds detected 
by NANTEN.  
The remaining 12 YSOs might be associated with small pc-sized molecular
clouds with masses lower than a few $\times10^4$ $M_\odot$, the NANTEN detection limit.
The high-resolution {\it HST} H$\alpha$ images show that indeed 7 of these
are in a visibly dusty environment.
These results indicate that at least 95\% of the 82 YSOs we examined 
are still associated with molecular material.

In Table~1 we summarize the YSOs and their properties: column 1 is the 
running number; column 2 lists the identifier of the YSOs from \citet{gruendl08}; 
column 3 describes the environment derived from {\it HST} H$\alpha$ images;
columns 4 and 5 give the names of the associated \ion{H}{2} region 
from \citet{davies76} and \citet{henize56}, respectively; 
column 6 lists whether the YSO is associated with molecular clouds 
detected by the NANTEN survey \citep{fukui01}; column 7 lists 
nearby OB associations \citep{lucke70}; columns 8 and 9 are described in
\S{5}; and the last column is described in \S{6}. In some cases YSOs are found
in more than one kind of environment, e.g., YSOs which are surrounded by
small \ion{H}{2} regions and are also associated with bright-rimmed dust 
pillars.  For such YSOs, column 3 indicates both the environments.  There
are also some cases where more than one YSO is discovered in the {\it Spitzer} 
PSF.  Such cases are indicated in column 3 as well.

\section{Properties of \ion{H}{2} Regions}

Of the 82 YSOs, 22 show small ionized regions that are well resolved, marginally 
resolved, or unresolved by the {\it HST} PSF (FWHM $\sim$ 0\farcs1).
For \ion{H}{2} regions with different degrees of resolution, different methods are
needed to measure the H$\alpha$ fluxes from the {\it HST} images and to assess whether
the unresolved sources possess H$\alpha$ line emission.

{\bf Resolved \ion{H}{2} Regions}--
For the eight YSOs that show well resolved \ion{H}{2} regions, we
measured the H$\alpha$ fluxes of the \ion{H}{2} regions using the {\it HST}
images, following the procedures for narrow-band WFPC2 
photometry\footnote{Available at http://www.stsci.edu/instruments/wfpc2/wfpc2\_faq/wfpc2\_nrw\_phot\_faq.html}.  
Images were divided by the exposure time and then the count rates were multiplied 
by the PHOTFLAM parameter found in the image headers to get flux densities.  
To obtain the fluxes, we multiplied the flux densities with the rectangular filter 
width calculated with SYNPHOT to be 28.3\AA\ and 74.9\AA\ for the filters F656N 
and F658N, respectively.  

To remove the stellar continuum from the integrated H$\alpha$ flux of a
well-resolved \ion{H}{2} region, we simply excised point sources from the 
H$\alpha$ image and replaced them with the average of the surrounding diffuse 
emission, as most of the \ion{H}{2} regions do not have broad-band continuum
images available.  One YSO in a resolved \ion{H}{2} region, 
YSO J052207.27-675819.7, has continuum images in the F675W band.
Using the F675W image and the H$\alpha$ image, we estimate that the stellar continuum 
contributes $\sim$17\% of the total observed H$\alpha$ flux 
within the boundary of the \ion{H}{2} region, while using the H$\alpha$ image alone we find
the stellar flux contributes $\sim$14\% of the total flux.
These results suggest that excising point sources from the H$\alpha$ images
is adequate for continuum subtraction.  The continuum-subtracted H$\alpha$ fluxes
of the \ion{H}{2} regions are listed in Table~2.


Assuming an electron temperature of 10$^4$ K, the H$\alpha$ surface brightness (SB)
of an \ion{H}{2} region can be expressed as
\begin{center}
$SB = 1.9 \times 10^{-18} EM$ ergs s$^{-1}$ cm$^{-2}$ arcsec$^{-2}$,\\
\end{center}
where $EM \equiv n_e^{2} L_{pc}$ is the emission measure of the \ion{H}{2} region; 
$n_e$ is the rms electron density in cm$^{-3}$, and $L_{pc}$ is the emitting path 
length in parsecs.  We measured the surface brightness of the \ion{H}{2} regions
directly from the images.  The variations in surface brightness over these
small \ion{H}{2} regions are not significantly large.  The peak brightnesses 
of the \ion{H}{2} regions are typically higher by a factor of 1.2--1.5 as 
compared to the average brightnesses of the \ion{H}{2} regions.  We used the 
average value of the surface brightness and determined the emission measure of 
each \ion{H}{2} region using the above relation.  From the emission measure and 
the average size of the \ion{H}{2} region (which we used as the average emitting
path length of these \ion{H}{2} regions), we then determined the rms electron 
density of the \ion{H}{2} region.  We also calculated extinction-uncorrected 
H$\alpha$ luminosities from the H$\alpha$ fluxes, and the required ionizing powers
Q(H$^0$) of the \ion{H}{2} regions for their H$\alpha$ luminosities.  Finally, we 
assessed the corresponding spectral types for the massive YSOs using the 
theoretical ionizing powers provided by \citet{panagia73}.   Table~2 lists the 
observed and derived properties of all the resolved \ion{H}{2} regions.
Note that the H$\alpha$ luminosities, ionizing powers, and the spectral types
given in Table~2 are lower limits to these quantities, as we have not applied an
extinction correction.


{\bf Marginally Resolved \ion{H}{2} Regions} --
There are four YSOs that show marginally resolved \ion{H}{2} regions.  
We did not find $UBV$ photometry for any of these four YSOs in 
the Magellanic Cloud Photometric Survey (MCPS; \citealt{zaritsky04}).  We searched 
for broad-band continuum WFPC2 and ACS images in the {\it HST} archive,
and found useful images for two YSOs, J052212.24-675813.1, and 
J053838.36-690630.4.  For J052212.24-675813.1, we found a continuum 
image in the wide-band filter F675W, and measured the continuum flux
of the YSO.  Scaling it according to the bandwidth, we find the expected
continuum flux in the H$\alpha$ band to contribute only 25\% of the total
flux measured.  The excess H$\alpha$ emission confirms the existence of
a small \ion{H}{2} region.  For the other YSO, J053838.36-690630.4, 
we found the continuum images in the wide-band filters F555W and F814W.  
We measured the fluxes of the YSO in these two bands and interpolated
between them to make a rough estimate of its expected continuum flux 
in the H$\alpha$ band.  The estimated continuum flux is  $\sim$21\% of 
the observed flux in the H$\alpha$ band.  This excess H$\alpha$ emission
also confirms the existence of a small \ion{H}{2} region.
The H$\alpha$ fluxes of these two marginally resolved \ion{H}{2} regions
(see Table 3) suggest that the central stars are early-type B stars, but 
cooler and less powerful than the earliest B stars seen in the resolved 
\ion{H}{2} regions in Table 2.  It is likely that the other two
YSOs with extended image but no photometric data are also in small
\ion{H}{2} regions, which need to be confirmed by spectroscopic 
observations in the future.

{\bf Unresolved \ion{H}{2} Regions} --
There are 10 YSOs that are unresolved in the {\it HST} H$\alpha$ images, but
it can be deduced from photometric analysis that they exhibit excess H$\alpha$ 
emission indicating the existence of unresolved \ion{H}{2} regions.
No useful {\it HST} continuum images are available for these YSOs, but we 
found MCPS $UBV$ photometry for seven of them and used these data to estimate
expected stellar continuum fluxes in the H$\alpha$ band.  We began by assuming
$(B-V)_0 \sim -0.3$, the color of O and early B main sequence stars, which 
have ionizing powers.  Adopting the canonical extinction relation $A_V \sim$
$3.2E(B-V)$, $E(U-B)/E(B-V) = 0.72$ for early-type stars, and a distance modulus
of 18.5, we calculated their absolute magnitudes in $UBV$ and the intrinsic 
color $(U-B)_0$, and compared these results with standard stars \citep{schmidt}
to determine the main sequence spectral types of the YSOs.  This approach works well, and
we can usually narrow down the classification to within 1--2 subtypes.  
Using the stellar effective temperatures implied by the spectral types, we then calculated
the expected (blackbody) continuum fluxes of the YSOs in the H$\alpha$ passband 
in the WFPC2 F656N or ACS F658N filter using the {\it calcphot} task in the SYNPHOT package.

For four YSOs, J045651.82-663133.0, J053549.28-660133.5, J053609.54-691805.5, 
and J054853.64-700320.2, their expected continuum fluxes in the H$\alpha$ 
passband are only 10--76\% of their observed H$\alpha$ fluxes.  If we take
into account the stellar photospheric absorption at the H$\alpha$ line,
the expected stellar fluxes are even lower in the H$\alpha$ passband.
Given that the equivalent widths of the H$\alpha$ absorption in early-B stars
range from $\sim$3.5 \AA\ in B0 stars to $\sim$5 \AA\ in B2 stars 
\citep{didelon82}, or 10--17\% for the 28.3 \AA\ width of the
F656N filter passband, we are confident that these
4 YSOs indeed have small \ion{H}{2} regions that are not resolved by 
the 0\farcs1 PSF of WFPC2. 

For three YSOs, J045720.72-662814.4, J053855.56-690426.5, and J053858.42-690434.7,
the observed H$\alpha$ fluxes are comparable to or somewhat lower than their expected 
continuum fluxes in the H$\alpha$ passband.  Considering the stellar photospheric
absorption of the H$\alpha$ line and the fact that the observed H$\alpha$
fluxes were not extinction-corrected, it is likely that these three YSOs
also have some excess H$\alpha$ emission and small unresolved \ion{H}{2} regions.
The existence of unresolved \ion{H}{2} regions for J045720.72-662814.4 and
J053858.42-690434.7 is further supported by the presence of IR spectral features that
originate from ionized and partially ionized gas, as discussed later in Section 5.  
For the remaining three YSOs, J045429.42-690936.9, J053821.10-690617.2, and 
J054844.29-700360.0, we did not find existing $UBV$ photometry or any {\it HST} 
broad-band images to estimate the continuum contribution in the H$\alpha$ 
passband; hence, we cannot determine whether these three YSOs have \ion{H}{2}
regions.  Based on the results of the other seven YSOs, we consider it  
likely that these three YSOs also have unresolved \ion{H}{2} regions. 
Table~3 lists the observed H$\alpha$ fluxes and the expected continuum fluxes 
in the H$\alpha$ passband of these ten YSOs with unresolved \ion{H}{2} regions.

Finally, we note that 10--18\% of B stars are known to be emission-line B stars
\citep{jaschek83}, and that we are not able to distinguish emission-line B stars
from unresolved \ion{H}{2} regions.  It is nevertheless certain that stellar
emission lines are formed in extended regions exterior to the photospheres,
and thus the distinction may be a matter of semantics.

\section{YSOs with {\it Spitzer} IRS Spectra}

\citet{seale09} presented {\it {\it {\it Spitzer}}} Infrared Spectrograph (IRS) 
spectra for 277 YSO candidates selected from the \citet{gruendl08} YSO catalog.  They found 
that the IRS spectra of massive YSO candidates can be divided into six different groups 
based on their spectral characteristics.   They proposed that these different 
groups reflect different evolutionary stages of massive YSOs.  The groups 
were defined as: spectra with silicate absorption, S group; spectra with 
silicate absorption and fine-structure lines, SE group; spectra showing Polycyclic 
Aromatic Hydrocarbons (PAH) emissions, P group; spectra showing both PAH emission 
and fine-structure lines, PE group, spectra showing fine-structure lines, E group; 
and finally spectra that appear to be featureless though they may show one or more of 
the above characteristics, F group.  \citet{seale09} noted that there were five YSOs
in their sample which they did not classify as 
the spectra for these five YSOs did not have full spectral coverage.  These 5 YSOs
showed fine-structure lines in the IRS spectra and were termed as
``embedded objects with unknown classification'' by \citet{seale09}.
Many of their P and PE group YSOs showed silicate absorption features at 10 $\mu$m.

Thirty-three of our YSOs with {\it HST} images have {\it Spitzer} IRS spectra 
reported by \citet{seale09}.  Columns 8 and 9 in Table~1 list the classification 
by \citet{seale09} of these YSOs based on the {\it Spitzer} IRS spectra, and
whether or not silicate absorption features are present in the IRS spectra, respectively.
The YSOs which were called as ``embedded objects with unknown classification''
are marked as ``Unknown'' in Table~1.
Eighteen YSOs in dark clouds have spectra, of which 14 are classified as PE 
or P group (where 9 of them show silicate absorption features), 2 are classified 
into the SE group, 1 is classified into the F group, and 1 is ``embedded object with 
unknown classification''.  Six YSOs in bright-rimmed 
globules have IRS spectra, of which, three are PE (with two of them in dark clouds 
showing silicate absorption features), one in a dark cloud is SE, and one is an 
``Embedded object with unknown classification''.   Among the 10 YSOs whose \ion{H}{2} regions are 
confirmed morphologically or photometrically, seven have IRS spectral type
PE (none with silicate features) and three are ``Embedded object with unknown 
classification''.  Finally, two YSOs whose \ion{H}{2} regions are not confirmed
either morphologically or photometrically, have IRS spectral type PE (no silicate 
absorption features) indicating that they have small 
unresolved \ion{H}{2} regions.

All the 12 YSOs with small \ion{H}{2} regions show fine-structure lines in the spectra
with 9 YSOs also showing PAH emission features.
This is understandable, as fine-structure lines 
are expected to be present in \ion{H}{2} regions of massive YSOs, whereas PAH 
emission features are generated in photodissociation regions on the surface of
\ion{H}{2} regions.  Many YSOs associated with dark clouds and bright-rimmed 
dust pillars also show PAH emission and/or fine-structure lines which implies 
that these YSOs might also have ionized regions surrounding them though not 
seen in the {\it HST} H$\alpha$ images.  It is interesting to note that none of the 
YSOs with small \ion{H}{2} regions show silicate absorption features, whereas 
half of the YSOs in dark clouds and one-third YSOs in bright-rimmed dust 
pillars show silicate absorption features in the IRS spectra.  Also,
two YSOs in dark clouds, one of which is associated with a bright-rimmed
dust pillar have been classified 
into the SE group---proposed by \citet{seale09} to be the less evolved stage compared to the P 
or PE group.  These results suggest that the YSOs in dark clouds and in bright 
rimmed dust pillars are less evolved compared to YSOs with small 
\ion{H}{2} regions. 

\section{YSO Categories and SEDs}
It is highly interesting that all the LMC YSOs for which we found archival {\it HST} 
data, are found in only one of three different kinds of environments.  Do the SEDs 
of these YSOs reflect any signatures that are characteristic of their environments?  
To check that, we examined the SEDs (optical to mid-infrared) of the YSOs and searched 
for any possible correlation between the YSO environments and their SEDs.  
We first classified YSOs according to the empirical ``Type'' classification 
based on the SEDs of massive YSOs, proposed by \citet{chen08}.  In this scheme, 
Type I has an SED that rises steeply from near-infrared to 24 $\mu$m and beyond,  
Type II has an SED with a low peak at optical wavelengths and a high peak at 
8--24 $\mu$m, and Type III shows an SED with bright optical peak and modest near- 
and mid-infrared peak (see Fig.~7 in \citealt{chen08}).  

We could classify 62 of our 82 YSOs using the above SED criteria (see the last column in Table~1).  
For the remaining YSOs, a classification could not be made 
either because there are multiple sources in the near-infrared data and/or in the 
{\it HST} data within the {\it Spitzer} PSF, making the YSO identification 
difficult, or the SED does not fit into any of the above Types, e.g., YSOs 
that are not detected in available optical data, and are faint or not detected 
at 24 $\mu m$.  Figure~2 shows the distribution of YSOs of different environments (column 3 in Table~1) 
into the Type I, II, and III classification (last column in Table~1).  For many YSOs, 
unambiguous classification into Type II or Type III was not possible; such cases are grouped as II/III.  

A correlation can be seen between YSO environments and their SEDs.  
YSOs in dark clouds with no optical counterparts are largely classified 
as Type I.  YSOs in dark clouds but with optical counterparts are mostly
classified as Type II.  YSOs in bright-rimmed dust pillars are classified 
as either Type II or classification is ambiguous between II and III.  
YSOs with resolved \ion{H}{2} regions are mostly classified as Type III, 
and YSOs with marginally resolved or unresolved \ion{H}{2} regions are classified as
Type II/III or Type III.  Thus, we see that YSOs in different environments 
are in different evolutionary stages.  YSOs in dark clouds are the youngest, 
YSOs with \ion{H}{2} regions are the most evolved, and YSOs in bright-rimmed 
dust pillars are in the intermediate stage.  Even among the YSOs in dark 
clouds, one can see a marked difference between the YSOs with and without 
optical counterparts.  The YSOs with optical counterparts are more evolved 
as compared to YSOs without optical counterparts.  The YSOs with marginally resolved and 
unresolved \ion{H}{2} regions are a mix of Type II and Type III, as opposed 
to YSOs with resolved \ion{H}{2} regions which are mostly Type III.  
This might mean that YSOs with marginally resolved and unresolved \ion{H}{2} regions are 
more likely to be younger compared to YSOs with resolved \ion{H}{2} regions.

We examined the {\it HST} H$\alpha$ images of YSOs to check for multiplicity.  
As many as 50\% YSOs show multiple sources within 2$\arcsec$ of the YSO location 
in the high-resolution {\it HST} images, of which 12\% show multiple YSOs.  
In the remaining cases, the other sources are normal stars.  Many of the 
multiples are also well resolved in the ISPI $J$ and $K_s$-bands. 
As has been cautioned by \citet{chen08}, multiplicity is a problem in 
interpreting the nature of LMC YSOs using only {\it Spitzer} data.  While 
the \citet{chen08} study is based only on one \ion{H}{2} complex, N44, the 
current study covers a wide range of star formation environments in the LMC
and demonstrates that multiplicity is indeed a prevailing problem.

\section{Triggered Star Formation}
The formation of massive stars has a significant impact on the structure and 
evolution of the interstellar medium.  After their birth, massive stars radiatively
ionize their ambient medium and mechanically energize their surroundings via
fast stellar winds and supernova ejecta.  While such energy feedback may
disperse the natal molecular cloud and terminate the star formation eventually,
the initial pressure increase in the ionization front on molecular material
may actually trigger star formation.  Using {\it HST} H$\alpha$ images, we
were able to examine the relationship between YSOs and ionization fronts
of antecedent massive stars.

Bright-rimmed dust pillars are potential sites of triggered star formation 
caused by compression due to ionization/shock fronts.  One fourth of our
sample, 19 YSOs, are found to be associated with 
bright-rimmed dust pillars.  One of these YSOs is found in a dark cloud
with an optical counterpart (2/1b) and four are found in dark clouds
with no optical counterparts (2/1a), indicative of their youth. 
Three bright-rimmed dust pillars show marginally resolved 
small \ion{H}{2} regions and harbor massive YSOs.  In order to assess 
whether the star formation was induced by external pressure in these
bright-rimmed dust pillars, we calculated thermal pressures of the ionized 
rims of the dust pillars, and present a few examples here.  

YSO J045659.85$-$662425.9 is associated with a bright-rimmed dust pillar that
shows the brightest ionized rims in all of our sample.  The peak H$\alpha$ 
surface brightness of the ionized gas enveloping this dust pillar is 
2.5 $\times$ 10$^{-12}$ ergs s$^{-1}$ cm$^{-2}$ arcsec$^{-2}$, corresponding 
to an emission measure of $1.31\times10^6$ cm$^{-6}$ pc, for a temperature 
of 10$^4$ K.  We adopt the average of the projected emission length of 
$\sim$0.55 pc and width of $\sim$0.18 pc (measured from H$\alpha$ image)
as the emission path length, 0.36 pc, and derive a rms electron density
of 1900 cm$^{-3}$.  The thermal pressure of the ionized rim of the dust 
globule is $P/k$ $\sim$1.9 $\times$ 10$^{7}$ cm$^{-3}$ K.  
The thermal pressure of the dust globule is $P/k$ $\sim$10$^{4}$ cm$^{-3}$ K, 
assuming typical values of density, 10$^3$ H$_2$ cm$^{-3}$, and temperature, 
10 K, for Bok globules.  The pressure of the ionized surface of the pillar 
is thus much higher than the thermal pressure of the dust globule, suggesting 
that star formation was possibly triggered by the external pressure.  
On another extreme, the YSO J045641.23$-$663132.9 is associated with a 
bright-rimmed dust pillar that shows the faintest rim among those in our sample.  
For this case, the peak H$\alpha$ surface brightness
of the ionized gas enveloping the dust pillar is 
3.7 $\times$ 10$^{-14}$ ergs s$^{-1}$ cm$^{-2}$ arcsec$^{-2}$, implying an 
emission measure of $1.95\times10^4$ cm$^{-6}$ pc.  The projected emission 
length and width on the surface of dust pillar are 0.8 and 0.6 pc, 
respectively, as measured from the H$\alpha$ image.  Using the average 
of these two values as the emission length along the line of sight,
we derive a rms electron density of $\sim$170 cm$^{-3}$, and the thermal 
pressure of the rim $P/k$ $\sim$1.7$\times$ 10$^{6}$ cm$^{-3}$ K.  
This is still much higher compared to the typical thermal pressure
of dust globules and could also be a case of triggered star formation.    
We therefore conclude that most of our YSOs in the bright-rimmed dust 
pillars have likely formed as a result of triggering due to external pressure.
    
Apart from the YSOs in the bright-rimmed dust pillars, there are 2 YSOs 
in dark clouds in juxtaposition with \ion{H}{2} regions, which may also
represent triggered star formation.  One of these two YSOs,
J053630.81-691817.2, is resolved into a stellar source and a small bright
\ion{H}{2} region separated by $\sim$0\farcs9 in the {\it HST} H$\alpha$ image.
The \ion{H}{2} region contains an unresolved point source near its peak
emission, indicating that the star might be its ionizing source.  The
ISPI $J$ and $K_s$ images also show two corresponding sources, with the
YSO being redder than the source in the \ion{H}{2} region.
The close proximity between the \ion{H}{2} region and the YSO, projected
distance $\sim$0.2 pc, suggests that the expansion of the \ion{H}{2}
region may have compressed the molecular cloud and triggered the formation
of the YSO.  Another such example is presented by the YSO
J045625.99-663155.5, which is projected at $\sim$1 pc from a resolved 
small \ion{H}{2} region, and is also a possible case of triggered star 
formation.  The high-resolution {\it HST} images are indeed 
very useful to examine the large scale star formation environments in detail,
and a systematic study of this nature is important to address issues related 
to formation of massive stars and the interplay of massive stars with the 
interstellar medium.

\section{Summary}

We have used archival {\it HST} H$\alpha$ images to examine the immediate
environments of massive and intermediate-mass YSO candidates of the LMC.  
In total, archival {\it HST} H$\alpha$ images were found for 99 YSO candidates.  
By examining the source morphology and the environment of YSO candidates in 
multiwavelength images, we conclude that 82 of the candidates are genuine YSOs.
All of these 82 YSOs are found in only three different environments, in dark 
clouds, in bright rimmed dust pillars, or in small \ion{H}{2} regions.  
Forty-nine YSOs are in dark clouds, 34 of which show no optical counterparts, and
the remaining 15 show faint optical counterparts in {\it HST} H$\alpha$ images. 
Nineteen YSOs are associated with bright-rimmed dust pillars, of which
five are in dark clouds, and three show small \ion{H}{2} regions.
Twenty two YSOs show small \ion{H}{2} regions, of which 8 are
well resolved, 4 are marginally resolved, and the remaining 10 are 
unresolved in the {\it HST} H$\alpha$ images.  We calculate observed (reddened)
H$\alpha$ fluxes of the resolved \ion{H}{2} regions and present
the estimated spectral types of these massive YSOs.  For the marginally resolved
as well as the unresolved \ion{H}{2} regions, we use 
$UBV$ photometry or {\it HST} continuum images to estimate the 
stellar continuum fluxes in the H$\alpha$ passband, and demonstrate that most
of them possess H$\alpha$ line emission, indicating that they indeed
have small \ion{H}{2} regions.

YSOs for which {\it Spitzer} IRS spectra are available, predominantly 
show PAH emission and/or fine-structure lines in the spectra.
Nine YSOs associated with dark clouds, with two of them also in
bright-rimmed dust pillars, show silicate absorption features
in the IRS spectra, whereas none of the YSOs associated with small 
\ion{H}{2} regions show silicate absorption features.
YSOs in small \ion{H}{2} regions including the marginally resolved,
and the unresolved ones, show PAH emission and/or fine-structure lines
in the IRS spectra.  The comparison of 
YSO environments with the SEDs reveal an evolutionary sequence of YSOs in 
different environments.  YSOs in dark clouds with no optical counterparts 
are mostly Type I and hence the youngest.  YSOs in dark clouds but with 
optical counterparts are mostly Type II and more evolved compared to YSOs 
with no optical counterparts.  YSOs in bright-rimmed dust pillars are
either Type II or Type II/III and are in the intermediate stage.  
YSOs with resolved \ion{H}{2} regions are mostly Type III and are the 
most evolved.  Finally, YSOs with marginally resolved \ion{H}{2} regions, 
and unresolved \ion{H}{2} regions are a mix of Type II and Type III, and 
should be on average younger compared to YSOs with resolved \ion{H}{2} 
regions.  As many as 50\% YSOs are resolved into multiple sources when 
seen in {\it HST} images, signifying the importance of using high-resolution 
images to probe the true nature of YSOs and to study their immediate 
environments.  

We investigate the issue of triggered star formation for YSOs in
bright-rimmed dust pillars and YSOs in dark clouds adjacent to 
\ion{H}{2} regions. The thermal pressures of ionized surfaces of bright-rimmed 
dust pillars are found to be much higher compared to typical thermal 
pressures of dust globules.  Thus the YSOs in bright-rimmed dust 
pillars have likely formed due to triggering from external pressure.  
Finally, we show that by examining the immediate environments of the YSOs
using the high-resolution {\it HST} images, we can learn about the evolutionary
stages of massive YSOs.  A systematic survey of massive YSOs using {\it HST}
will be very useful to study the evolutionary aspects of massive YSOs and to 
understand star formation in wide range of interstellar environments. 

\acknowledgements 
This work was supported through NASA grants HST-AR-10942.01-A, JPL 1290956,  
and JPL 1316421.  This work made use of 
the data products of the Two Micron All Sky Survey (2MASS), which is a joint project 
of the University of Massachusetts and the Infrared Processing and Analysis 
Center/California Institute of Technology, funded by the National Aeronautics and 
the Space Administration and the National Science Foundation.


\clearpage

\begin{deluxetable}{lccccccccc}
\tabletypesize{\scriptsize}
\setlength{\tabcolsep}{0.08in}
\tablecolumns{10}
\tablewidth{500pt}
\tablecaption{Properties of  YSOs with {\it HST} H$\alpha$ Images}
\tablehead{
\colhead{Number} & \colhead{ID\tablenotemark{a}} & \colhead{Category\tablenotemark{b}} & \colhead{H~{\sc ii}} & 
\colhead{H~{\sc ii}} & \colhead{CO} & \colhead{OB} &\colhead{IRS} & \colhead{Silicate} & 
\colhead{YSO} \\
\colhead{~} & \colhead{~} & \colhead{~} & \colhead{Region} & \colhead{Region} & \colhead{detection} & \colhead {Association} 
& \colhead{Spectra} & \colhead{Absorption} & \colhead{Type}}
\startdata
\cutinhead{YSOs in Dark Clouds}
1& J045625.99-663155.5& 1a+1a\tablenotemark{\ast} & DEM34& N11F& Yes& LH9 & PE & No & I\\
2& J045629.02-663159.3& 1a& DEM34& N11F& Yes& LH9 & PE & No & II\\
3& J045640.79-663230.5& 1a& DEM34& N11F& Yes& LH9 & F & No & I \\
4& J045742.00-662634.4& 1a& DEM34& N11C& Yes& LH13 & PE & No & I\\
5& J045747.68-662816.9& 1a& DEM34& N11D& Yes& LH13 & PE & Yes & I\\
6& J052207.32-675826.8& 1a& DEM152& N44C& Yes& LH47 & -- &-- & I\\
7& J052211.86-675818.1& 1a& DEM152& N44C& Yes& LH47 & -- & --& I\\
8& J052212.57-675832.4& 1a& DEM152& N44C& Yes& LH47 & SE & Yes & I\\
9& J053628.51-691636.6& 1a& DEM263& N157& No & --& -- & --& I\\
10& J053822.43-690644.4& 1a& DEM263& N157& Yes&  --& -- & -- & I\\
11& J053833.09-690611.9& 1a& DEM263& N157& Yes& LH100 & -- & -- & I\\
12& J053834.06-690452.2& 1a& DEM263& N157& Yes& LH100 & -- & --& I\\
13& J053834.60-690557.0& 1a& DEM263& N157& Yes& LH100 & -- & -- & I\\
14& J053843.52-690629.0& 1a+1b\tablenotemark{\ast}& DEM263& N157& Yes& LH100 & -- & -- & I\\
15& J053844.32-690329.9& 1a+1b\tablenotemark{\ast}& DEM263& N157& Yes& LH100 & -- & -- & II\\
16& J053848.17-690411.7& 1a& DEM263& N157& Yes& LH100 & -- & -- & I\\
17& J053848.42-690441.6& 1a& DEM263& N157& Yes& LH100 & -- & -- & I\\
18& J053849.27-690444.4& 1a& DEM263& N157& Yes& LH100 & PE & Yes & I\\
19& J053852.67-690437.5& 1a& DEM263& N157& Yes& LH100 & PE & No & II/III\\
20& J053909.08-693005.7& 1a& DEM269& N158C& Yes& LH101 & -- & -- & I\\
21& J053912.79-693041.8& 1a& DEM269& N158C& Yes& LH101 & -- & -- & I\\
22& J053922.74-693011.7& 1a& DEM269& N158C& Yes& LH101 & -- & -- & II\\
23& J053938.73-693904.3& 1a& DEM284& N160D& Yes& LH103 & PE & Yes & I\\
24& J053943.60-693820.4& 1a& DEM284& N160A& Yes& LH103 & -- & --& --\\
25& J053944.33-693847.5& 1a+1b+1b\tablenotemark{\ast}& DEM284& N160A& Yes& LH103 & -- & --&--\\
26& J054003.49-694355.0& 1a& DEM271& N159D& Yes& LH105 & -- & --&--\\
27& J054009.49-694453.5& 1a& DEM271& N159B& Yes& LH105 & PE & Yes & I\\
28& J054019.01-694445.6& 1a& DEM271& N159G& Yes& LH105 & -- & --&--\\
29& J054839.05-700536.4& 1a& DEM323& N180B& Yes& LH117 & -- & --&--\\
30& J045627.37-663201.0& 1b& DEM34& N11F& Yes& LH9 & -- & --& II\\
31& J045633.02-662359.8& 1b& DEM34& N11B& Yes& LH10 & -- & -- & --\\
32& J045638.76-662446.2& 1b& DEM34& N11B& Yes& LH10 & PE & No & I\\
33& J045725.75-662545.8& 1b& DEM34& N11& Yes&  -- & -- & --& II\\
34& J051351.51-672721.9& 1b& DEM106& N30B&  No & LH38 & P & Yes & II/III\\
35& J052208.52-675922.1& 1b& DEM152& N44C& Yes& LH47 & -- & --& II\\
36& J053630.81-691817.2& 1b& DEM263& N157& No & --& P& No &II\\
37& J053838.45-690418.3& 1b& DEM263& N157& Yes& LH100 & -- & --& II\\
38& J053904.88-692949.9& 1b& DEM269& N158C& Yes& LH101 & PE & No & II\\
39& J053906.31-693043.8& 1b& DEM269& N158C& Yes& LH101 & PE & Yes & II\\
\cutinhead{YSOs in Bright-Rimmed Dust Pillars}
40& J045628.68-663143.4& 2/1b\tablenotemark{\dagger}& DEM34& N11F& Yes& LH9 & -- & --& II\\
41& J045641.23-663132.9& 2& DEM34& N11F& Yes& LH9 & -- & --& II\\
42& J045657.25-662513.0& 2/1a\tablenotemark{\dagger}& DEM34& N11B& Yes& LH10 & -- & -- & II/III\\
43& J045659.85-662425.9& 2& DEM34& N11B& Yes& LH10 & PE & No & III\\
44& J045737.61-662726.7& 2& DEM34& N11C& Yes& LH13 & -- & --& III\\
45& J045739.04-662731.3& 2& DEM34& N11C& Yes& LH13 & -- & --&--\\
46& J051829.16-691458.1& 2& DEM132A& N119&  No & LH41 & -- & -- & II/III\\
47& J052135.83-675443.0& 2& DEM152A& N44F& Yes& LH47 & -- & --&--\\
48& J052601.20-673012.1& 2/1a\tablenotemark{\dagger}& DEM192& N51D& No & LH54 & SE & Yes&--\\
49& J052619.79-673033.6& 2& DEM192& N51D&  No & LH54 & -- & --&--\\
50& J053609.54-691805.5& 2/3c\tablenotemark{\dagger}& DEM263& N157& No & --& -- & --& II/III\\
51& J053623.52-691002.1& 2& DEM263& N157& No & --& -- & --& II/III\\
52& J053838.36-690630.4& 2/3b\tablenotemark{\dagger}& DEM263& N157& Yes& LH100 & -- & -- & III\\
53& J053839.23-690552.2& 2& DEM263& N157& Yes& LH100 & PE & No & II\\
54& J053839.69-690538.2& 2/1a\tablenotemark{\dagger}& DEM263& N157& Yes& LH100 & PE & Yes & II\\
55& J053912.67-692941.4& 2& DEM269& N158C& Yes& LH101 & -- & --&--\\
56& J053943.82-693834.0& 2/3b+2/1a\tablenotemark{\dagger,\ast} & DEM284& N160A& Yes& LH103 & Unknown & No &--\\
57& J054841.29-700536.7& 2& DEM323& N180B& Yes& LH117 & -- & -- & II/III\\
\cutinhead{YSOs in H~{\sc ii} regions}
58& J045426.06-691102.3& 3a& DEM22& N83B& Yes& LH5 & PE & No & III\\
59& J045708.84-662325.1& 3a& DEM34& N11A& Yes& LH10 & -- & --& III\\
60& J045716.25-662319.9& 3a& DEM34& N11& Yes& LH10 & PE & No & II/III\\
61& J052207.27-675819.7& 3a& DEM152& N44C& Yes& LH47 & PE & No & II/III\\
62& J053845.15-690507.9& 3a& DEM263& N157& Yes& LH100 & PE & No & III\\
63& J053943.26-693854.6& 3a& DEM284& N160A& Yes& LH103 & Unknown & No & III\\
64& J053945.94-693839.2& 3a& DEM284& N160A& Yes& LH103 & Unknown & No & II/III\\
65& J054004.40-694437.6& 3a& DEM271& N159B& Yes& LH105 & PE & No & III\\
66& J052212.24-675813.1& 3b& DEM152& N44C& Yes& LH47 & -- & --&--\\
67& J052249.13-680129.1& 3b& DEM160& N44H& Yes& LH49 & PE & No & II/III\\
68& J045429.42-690936.9& 3c& DEM22& N83& Yes& LH5 & -- & --&II\\
69& J045651.82-663133.0& 3c& DEM34& N11F& No & LH9 & PE & No & II\\
70& J045720.72-662814.4& 3c& DEM34& N11& No& -- & PE & No & II/III\\
71& J053549.28-660133.5& 3c& DEM243& N63&  No & LH83 & -- & --&--\\
72& J053821.10-690617.2& 3c& DEM263& N157& Yes&  --& --&-- & III\\
73& J053855.56-690426.5& 3c& DEM263& N157& Yes& LH100 & -- & -- & III\\
74& J053858.42-690434.7& 3c& DEM263& N157& Yes& LH100 & PE & No & III\\
75& J054844.29-700360.0& 3c& DEM323& N180B& Yes& LH117 & -- & -- & II/III\\
76& J054853.64-700320.2& 3c& DEM323& N180B& Yes& LH117 & -- & --& II/III\\
\enddata
\tablenotetext{a}{The identifier of the YSO shows the J2000 coordinates of the YSOs in {\it hhmmss.s-ddmmss.s} format.}
\tablenotetext{b}{Category describes YSO environments: YSOs found in dark clouds without optical counterparts
in {\it HST} H$\alpha$ images are 1a, YSOs in dark clouds with optical counterparts are 1b; YSOs in bright rimmed
dust pillars are 2; YSOs with resolved H~{\sc ii} regions are 3a, YSOs with
marginally resolved H~{\sc ii} regions are 3b, and YSOs with unresolved H~{\sc ii} regions are 3c.}
\tablenotetext{\dagger}{YSOs are found in more than one kind of environment, e.g. 2/3b refers to YSOs which
are associated with bright-rimmed dust pillars and the optical counterpart is a small H~{\sc ii} region.}
\tablenotetext{\ast}{More than one YSOs is discovered in the {\it Spitzer} PSF.}
\end{deluxetable}

\begin{deluxetable}{ccccccccc}
\tabletypesize{\scriptsize}
\setlength{\tabcolsep}{0.06in}
\tablecolumns{10}
\tablewidth{0pt}
\tablecaption{Properties of Resolved \ion{H}{2} Regions Associated with Massive YSOs}
\tablehead{
\colhead{ID} & \colhead{Length\tablenotemark{a}} &\colhead{H$\alpha$ Flux\tablenotemark{b}} & \colhead{EM} &
\colhead{$n_e$} &\colhead{$L_{H\alpha}$\tablenotemark{b}} &\colhead{$Q(H_0)$} &
\colhead{Sp.Type} & \colhead{IRS}\\
\colhead{~} & \colhead{(pc)}  & \colhead{(ergs cm$^{-2}$ s$^{-1}$)} & \colhead{(cm$^{-6}$ pc)} & \colhead{(cm$^{-3}$)} & \colhead{(ergs s$^{-1}$)} &
\colhead{(photons s$^{-1}$)} & \colhead{~} & \colhead{Spectra}}
\startdata
J045426.06-691102.3 & 0.96 & 6.84$\times$10$^{-12}$ &  2.86$\times$10$^{5}$ & 546 & 1.45$\times$10$^{35}$ & 1.03$\times$10$^{47}$ & B0.5 & PE \\
J045708.84-662325.1 & 1.80 & 2.55$\times$10$^{-12}$ & 7.59$\times$10$^{3}$ & 64 & 5.40$\times$10$^{34}$ & 3.94$\times$10$^{46}$ & B0.5 & --\\  
J045716.25-662319.9 &  1.68 & 8.64$\times$10$^{-12}$ &  1.18$\times$10$^{5}$ & 265  & 1.83$\times$10$^{35}$ & 1.35$\times$10$^{47}$ & B0.5 & PE \\
J052207.27-675819.7 & 0.33 & 3.19$\times$10$^{-14}$ &  1.08$\times$10$^{4}$ & 181 & 6.76$\times$10$^{32}$ & 4.80$\times$10$^{44}$ & B2 & PE \\
J053845.15-690507.9 & 0.81 & 2.38$\times$10$^{-12}$ &  1.37$\times$10$^{5}$ & 412 & 5.04$\times$10$^{34}$ & 3.59$\times$10$^{46}$ & B0.5 & PE \\
J053943.26-693854.6 & 0.96 & 7.92$\times$10$^{-12}$ &  3.33$\times$10$^{5}$ & 587 & 1.67$\times$10$^{35}$ & 1.19$\times$10$^{47}$ & B0 & Unknown \\
J053945.94-693839.2 & 1.44 & 1.07$\times$10$^{-11}$ &  1.98$\times$10$^{5}$ & 371  & 2.26$\times$10$^{35}$ & 1.61$\times$10$^{47}$ & B0 & Unknown \\
J054004.40-694437.6 & 0.96 & 2.78$\times$10$^{-12}$ &  1.16$\times$10$^{5}$ & 348 & 5.89$\times$10$^{34}$ & 4.20$\times$10$^{46}$ & B0.5 & PE \\
\enddata
\tablenotetext{a}{Average emitting path lengths of the H~{\sc ii} regions.}
\tablenotetext{b}{Not corrected for extinction.}
\end{deluxetable}

\clearpage
\begin{deluxetable}{ccccc}
\tabletypesize{\scriptsize}
\setlength{\tabcolsep}{0.06in}
\tablecolumns{5}
\tablewidth{0pt}
\tablecaption{Properties of Marginally Resolved and Unresolved \ion{H}{2} Regions Associated with Massive YSOs}
\tablehead{
\colhead{ID} & \colhead{Category} &\colhead{Observed H$\alpha$ Flux} & \colhead{Expected Cont. Flux\tablenotemark{a}}
 & \colhead{IRS}\\
\colhead{~} & \colhead{~} & \colhead{(ergs cm$^{-2}$ s$^{-1}$)} & \colhead{(ergs cm$^{-2}$ s$^{-1}$)} & \colhead{Spectra}}
\startdata
J052212.24-675813.1 & 3b  & 5.1 $\times$ 10$^{-15}$ & 1.3 $\times$ 10$^{-15}$ & -- \\
J052249.13-680129.1 & 3b  & 1.3 $\times$ 10$^{-15}$ & -- & PE \\
J053838.36-690630.4 & 3b  & 3.1 $\times$ 10$^{-13}$ & 6.7 $\times$ 10$^{-14}$ & --\\
J053943.80-693834.0 & 3b  & 3.0 $\times$ 10$^{-13}$ & -- & Unknown\\
\hline 
J045651.82-663133.0 & 3c  & 2.4 $\times$ 10$^{-14}$ & 1.8 $\times$ 10$^{-14}$  & PE \\
J045429.42-690936.9 & 3c  & 1.1 $\times$ 10$^{-13}$ & -- & --\\
J045720.72-662814.4 & 3c  & 6.7 $\times$ 10$^{-15}$ & 1.1 $\times$ 10$^{-14}$ & PE \\
J053549.28-660133.5 & 3c  & 7.0 $\times$ 10$^{-15}$ & 5.0 $\times$ 10$^{-15}$ & --\\
J053609.54-691805.5 & 3c  & 4.4 $\times$ 10$^{-14}$ & 1.0 $\times$ 10$^{-14}$ & --\\
J053821.10-690617.2 & 3c  & 3.1 $\times$ 10$^{-14}$ & -- & --\\
J053855.56-690426.5 & 3c  & 9.3 $\times$ 10$^{-15}$ & 1.0 $\times$ 10$^{-14}$ & -- \\
J053858.42-690434.7 & 3c  & 8.9 $\times$ 10$^{-15}$ & 9.7 $\times$ 10$^{-15}$ & PE\\
J054844.29-700360.0 & 3c  & 6.6 $\times$ 10$^{-15}$ & -- & --\\
J054853.64-700320.2 & 3c  & 3.4 $\times$ 10$^{-14}$ & 4.0 $\times$ 10$^{-15}$ & -- \\
\enddata
\tablenotetext{a}{Expected continuum flux of the YSO in the H$\alpha$ passband.}
\end{deluxetable}

\clearpage

\begin{figure}
\includegraphics[scale=0.6]{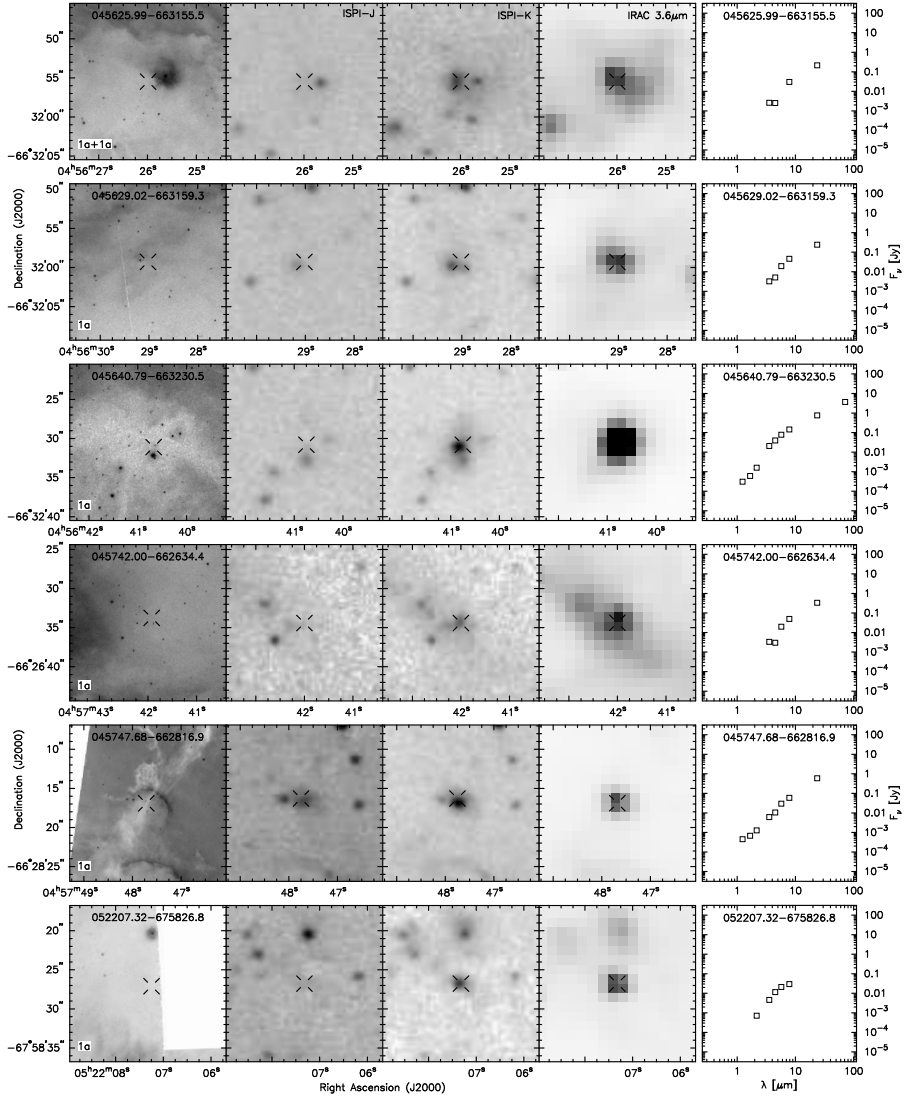}
\caption{From left to right each set of five panels show
  20\arcsec$\times$20\arcsec\ \emph{HST} H$\alpha$, ISPI $J$, $K_s$, and
  \emph{Spitzer} 3.6~$\mu$m images centered on the \emph{Spitzer} position
  for each YSO, and the SED of the YSO.  The position of the YSO and its optical near-IR
  counterparts are indicated with an open cross.  The category for each YSO is
  indicated in the lower left corner where: YSOs in dark clouds without
  and with optical counterparts are marked as categories 1a and 1b,
  respectively; YSOs in bright-rimmed dust pillars are marked as category 2;
  YSOs associated with well-resolved, marginally resolved, and unresolved
  H~{\sc ii} regions are marked as categories 3a, 3b, and 3c, respectively. 
  Cases where multiple optical/near-IR counterparts are discovered
  within the \emph{Spitzer} PSF are indicated with a "+" and 
  YSOs found in more than one kind of environment are indicated with a
  "/".} 

\end{figure}

\begin{center}
\includegraphics[scale=0.85]{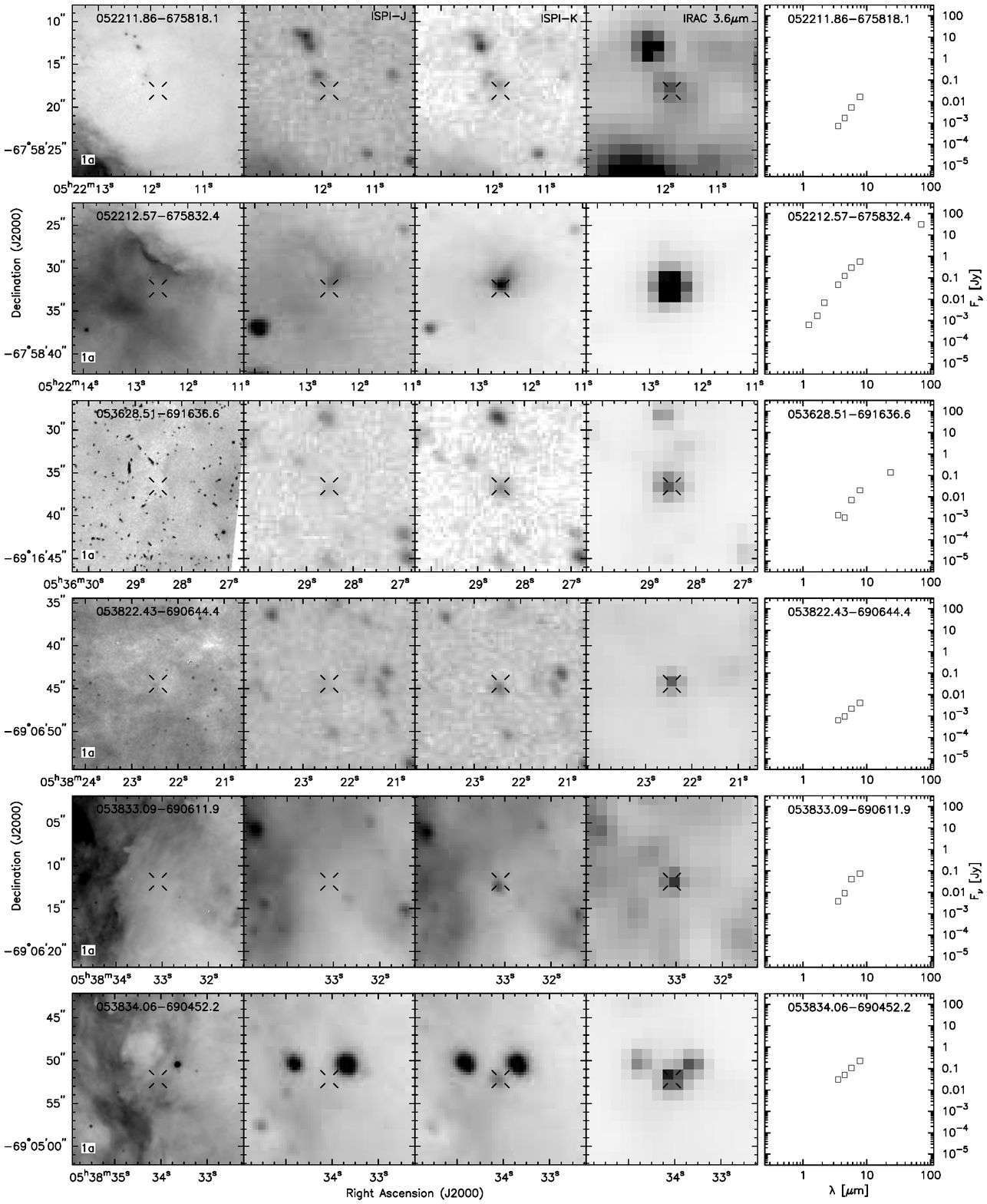}
\end{center}

\includegraphics[scale=0.85]{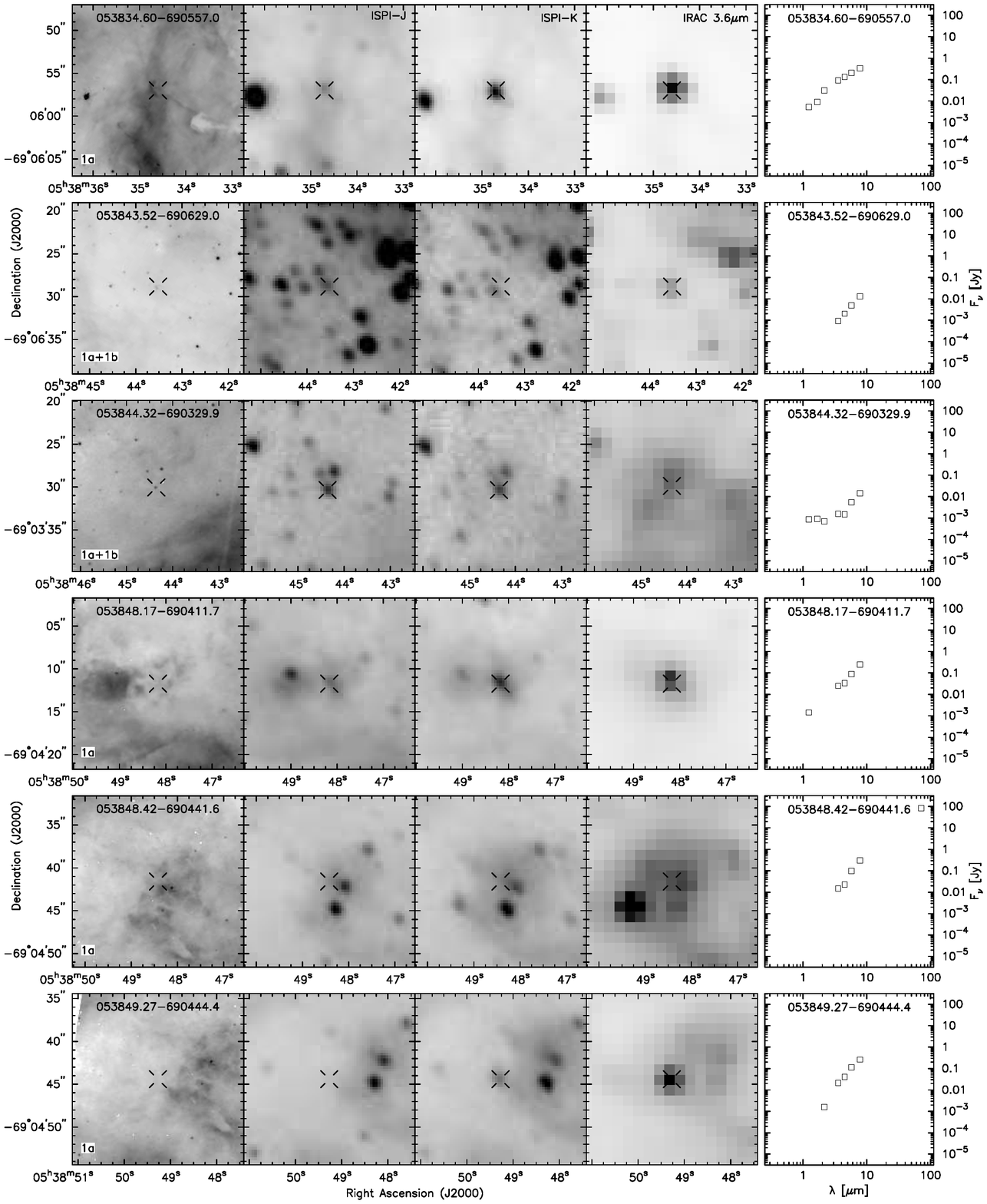}

\includegraphics[scale=0.85]{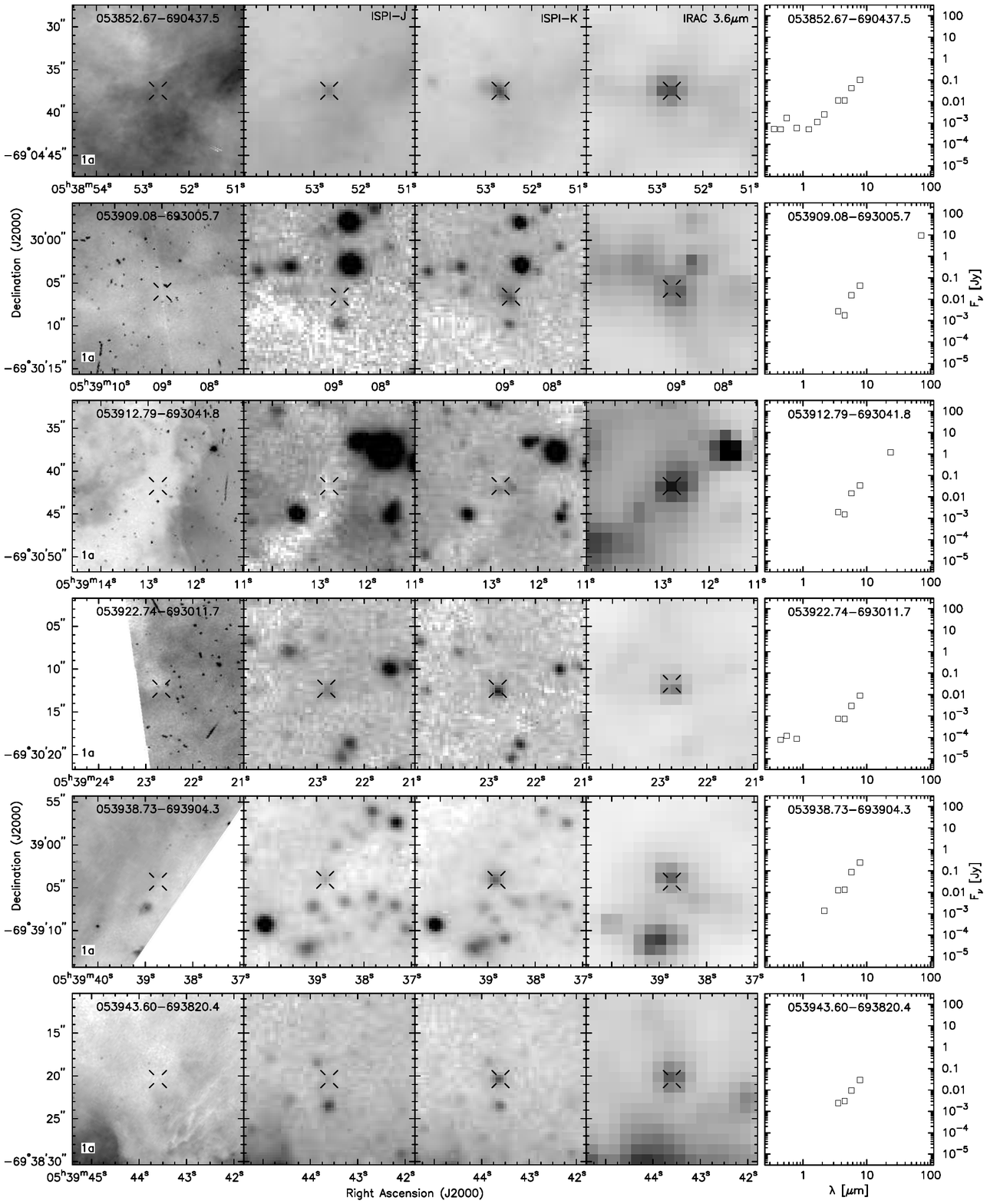}

\includegraphics[scale=0.85]{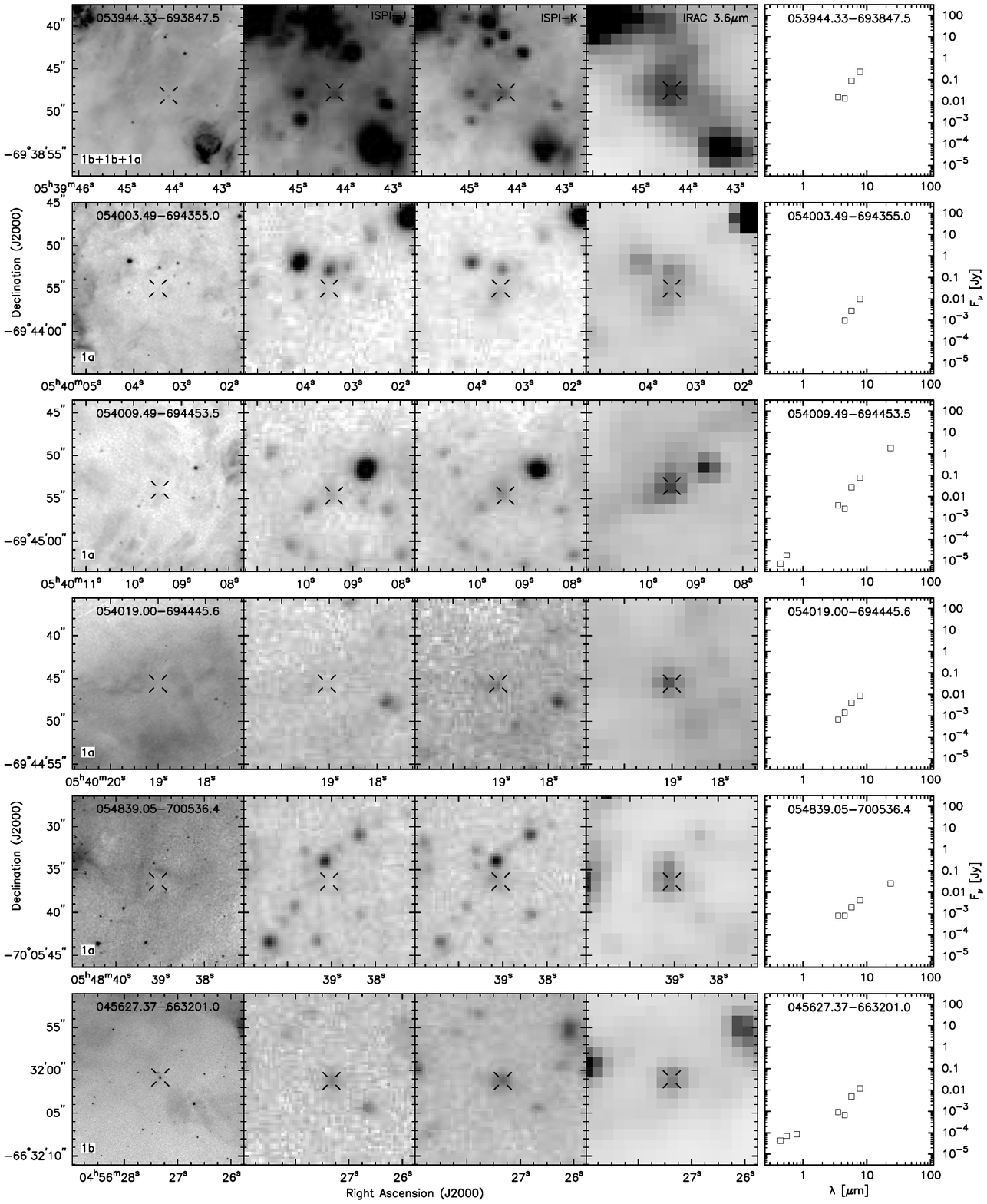}

\includegraphics[scale=0.85]{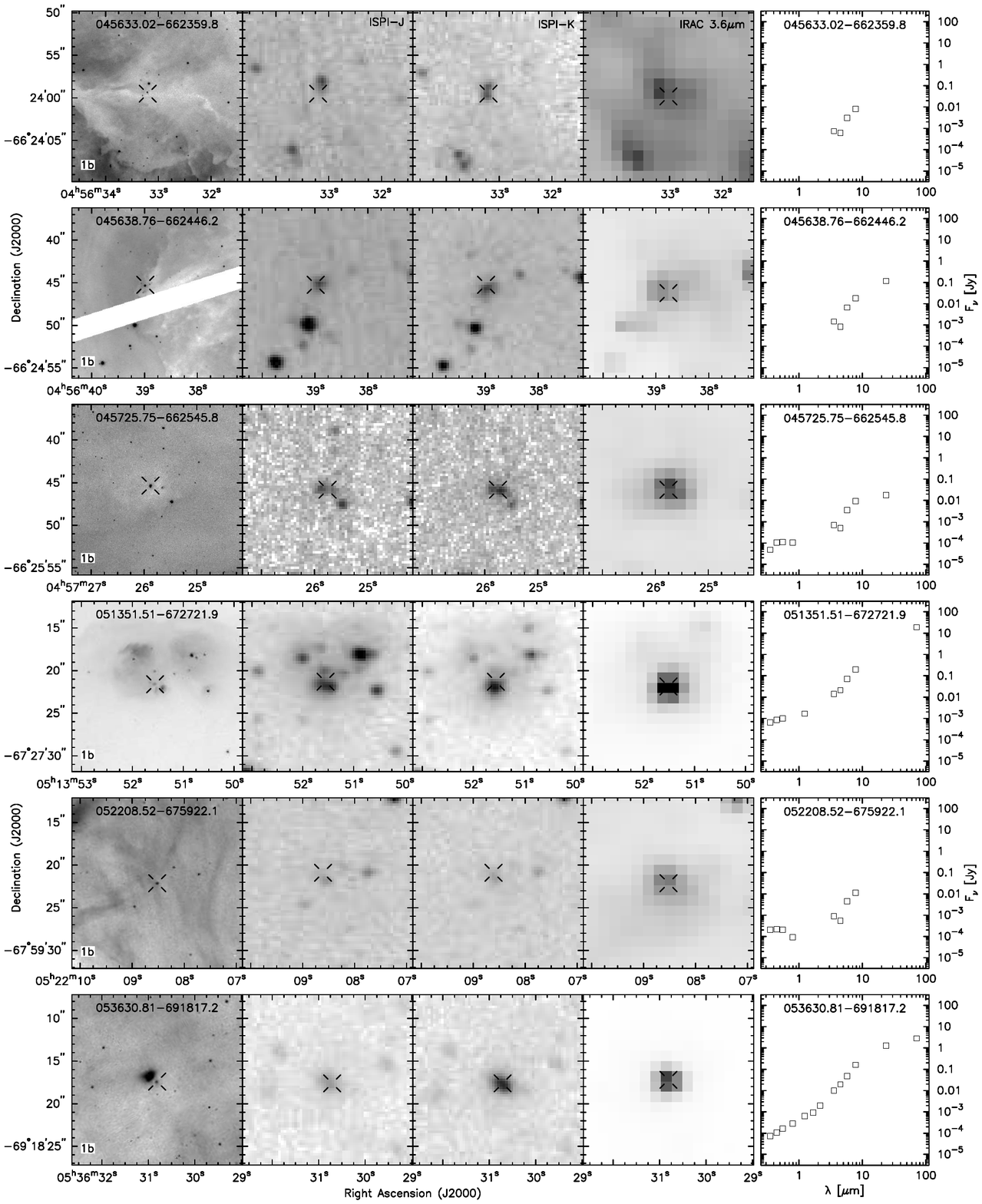}

\includegraphics[scale=0.85]{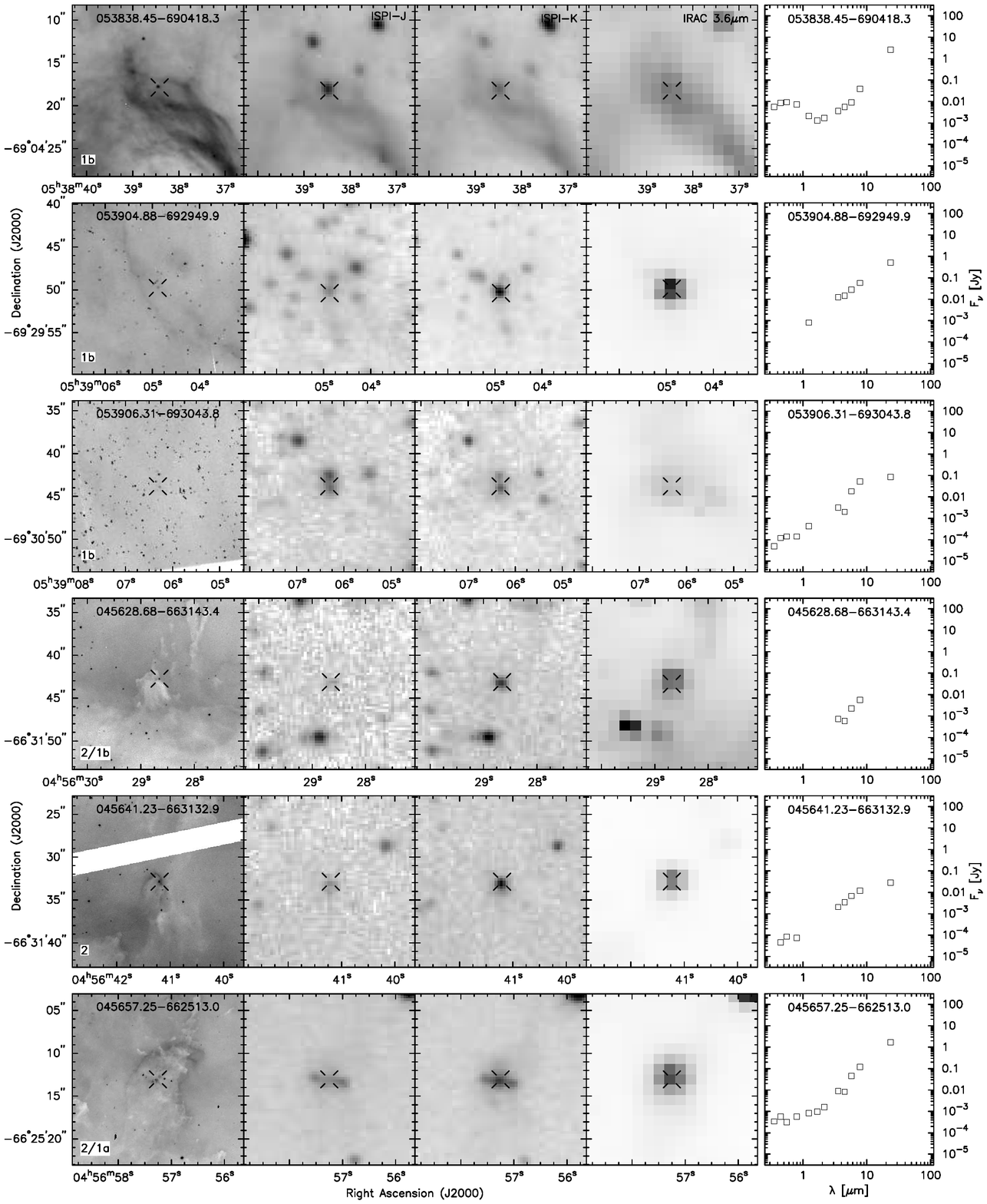}

\includegraphics[scale=0.85]{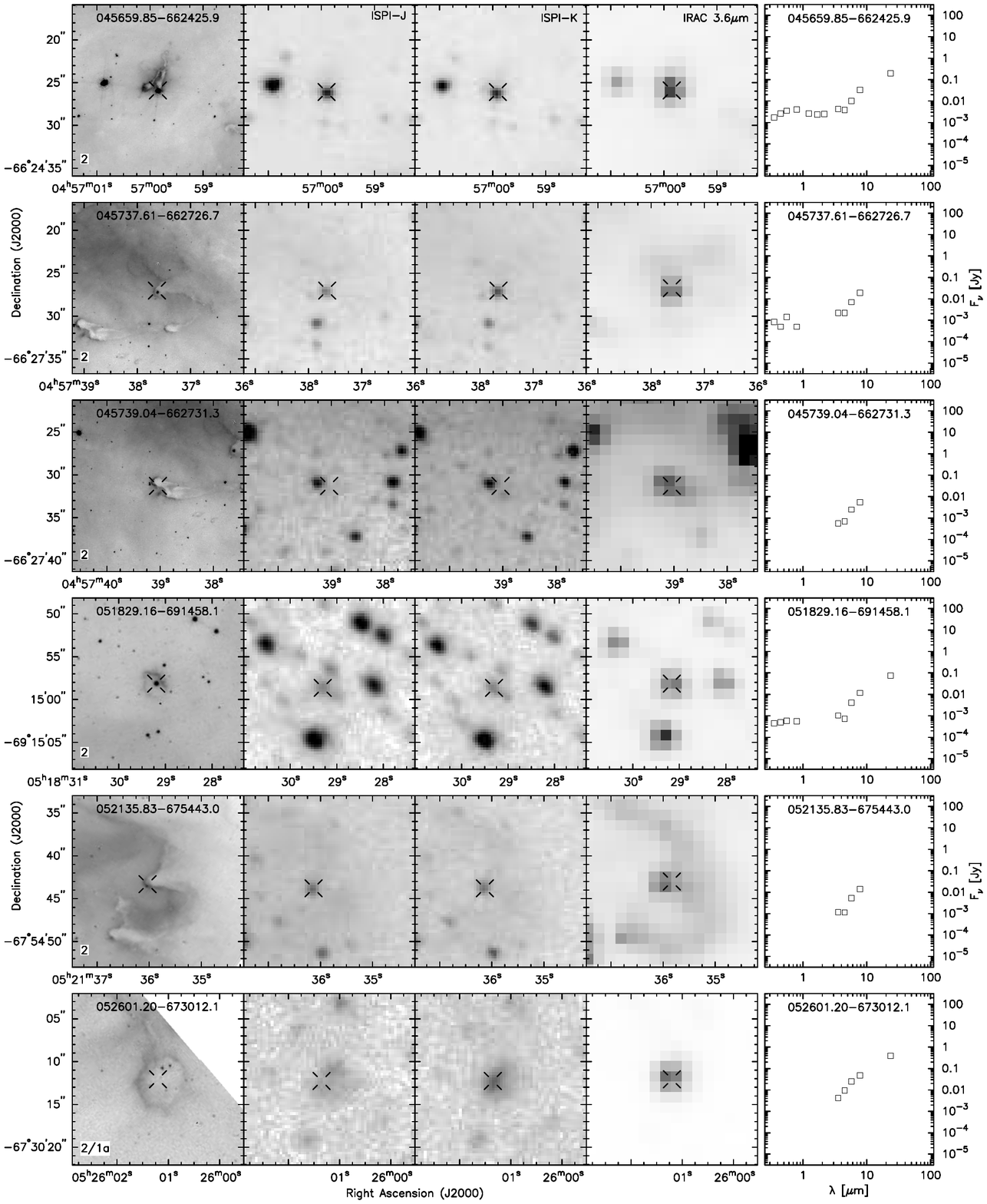}

\includegraphics[scale=0.85]{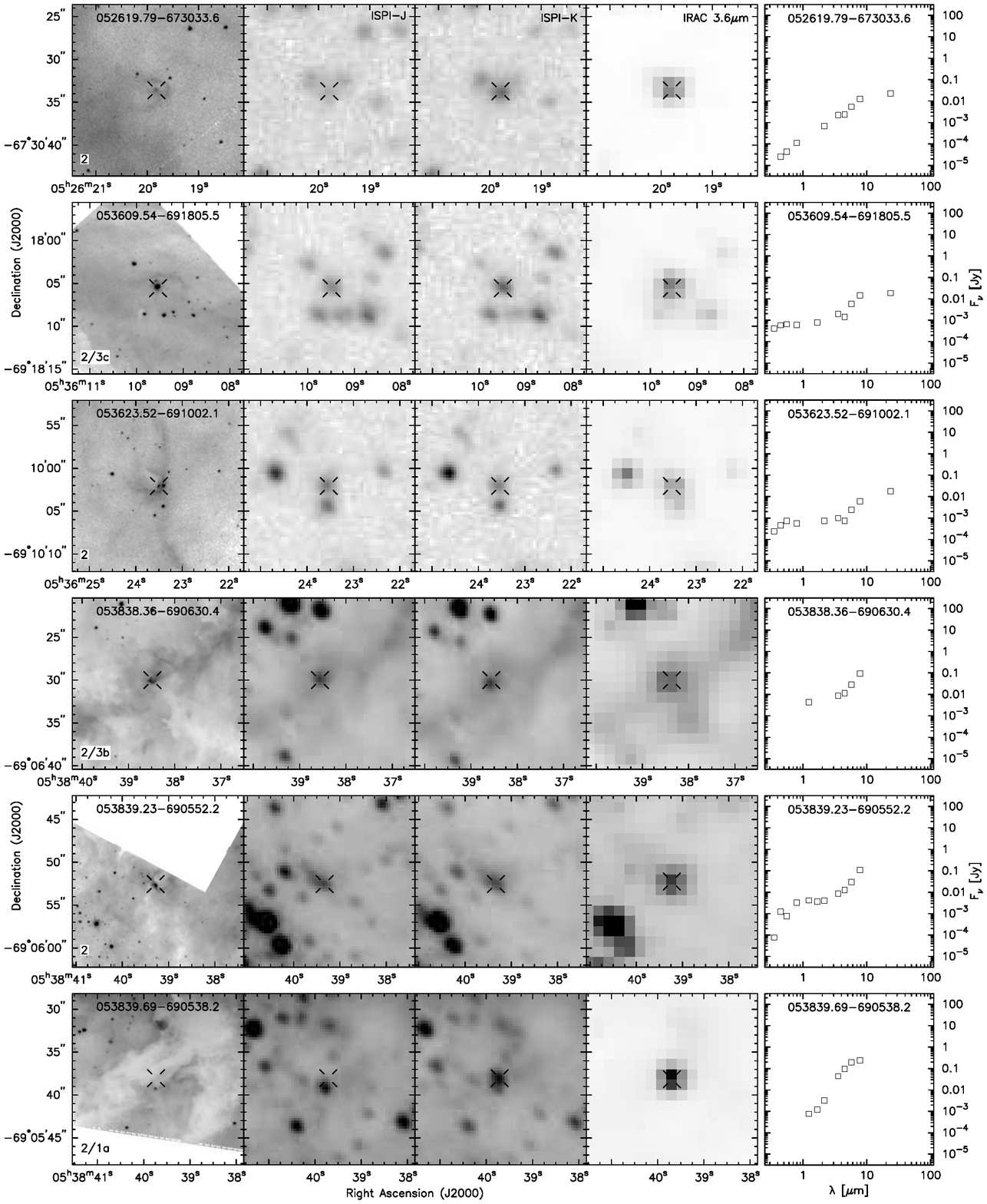}

\includegraphics[scale=0.85]{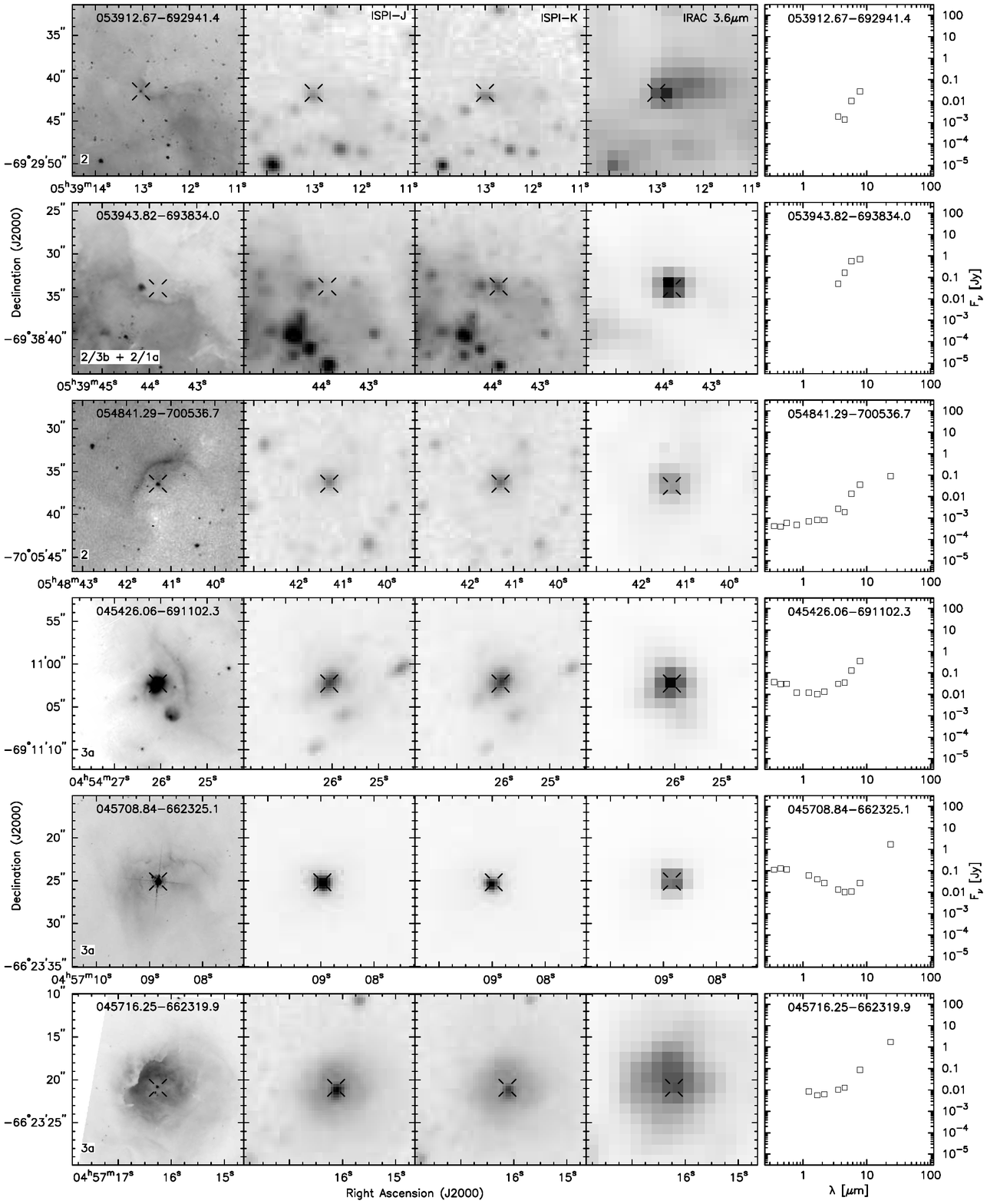}

\includegraphics[scale=0.85]{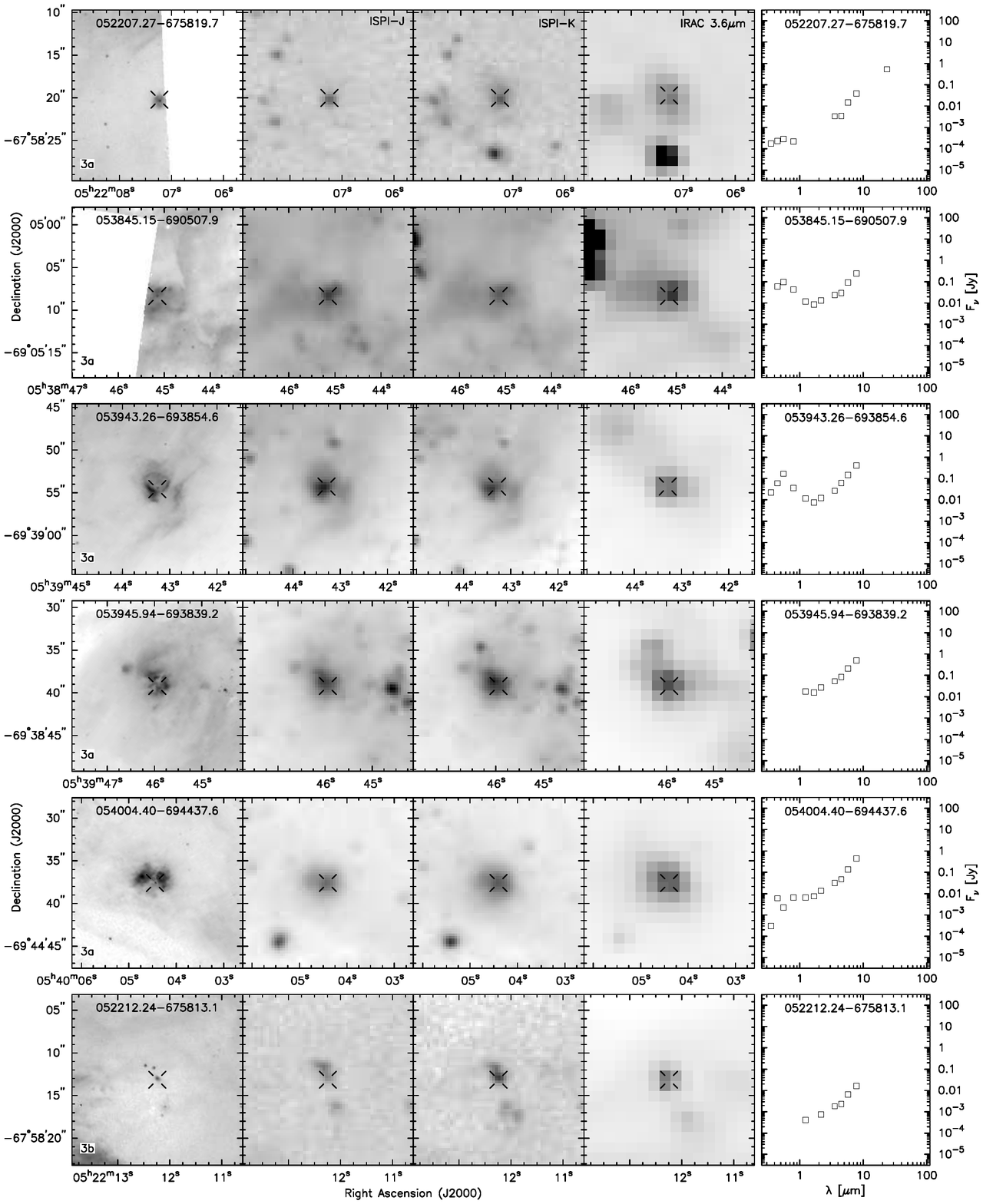}

\includegraphics[scale=0.85]{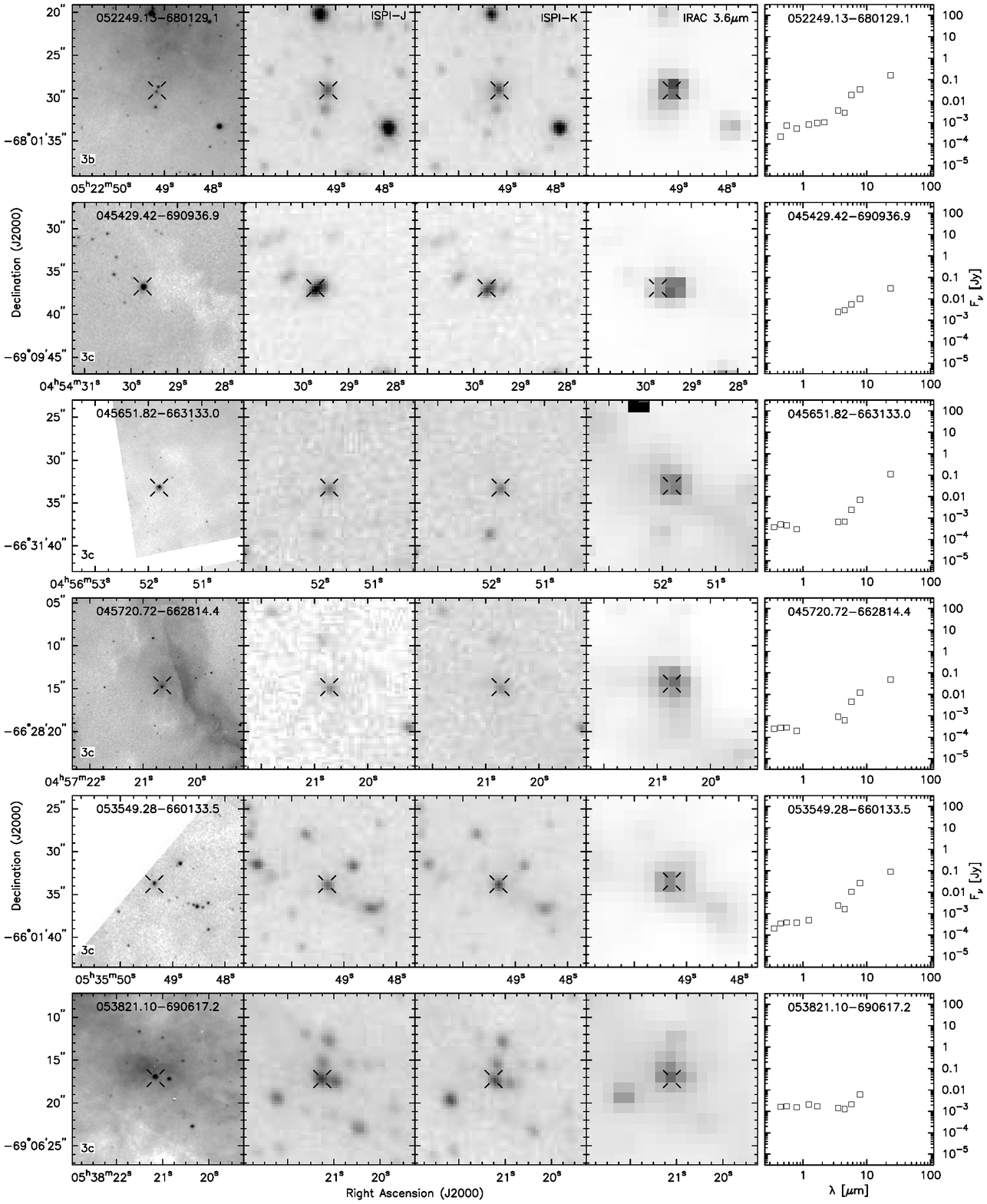}

\includegraphics[scale=0.85]{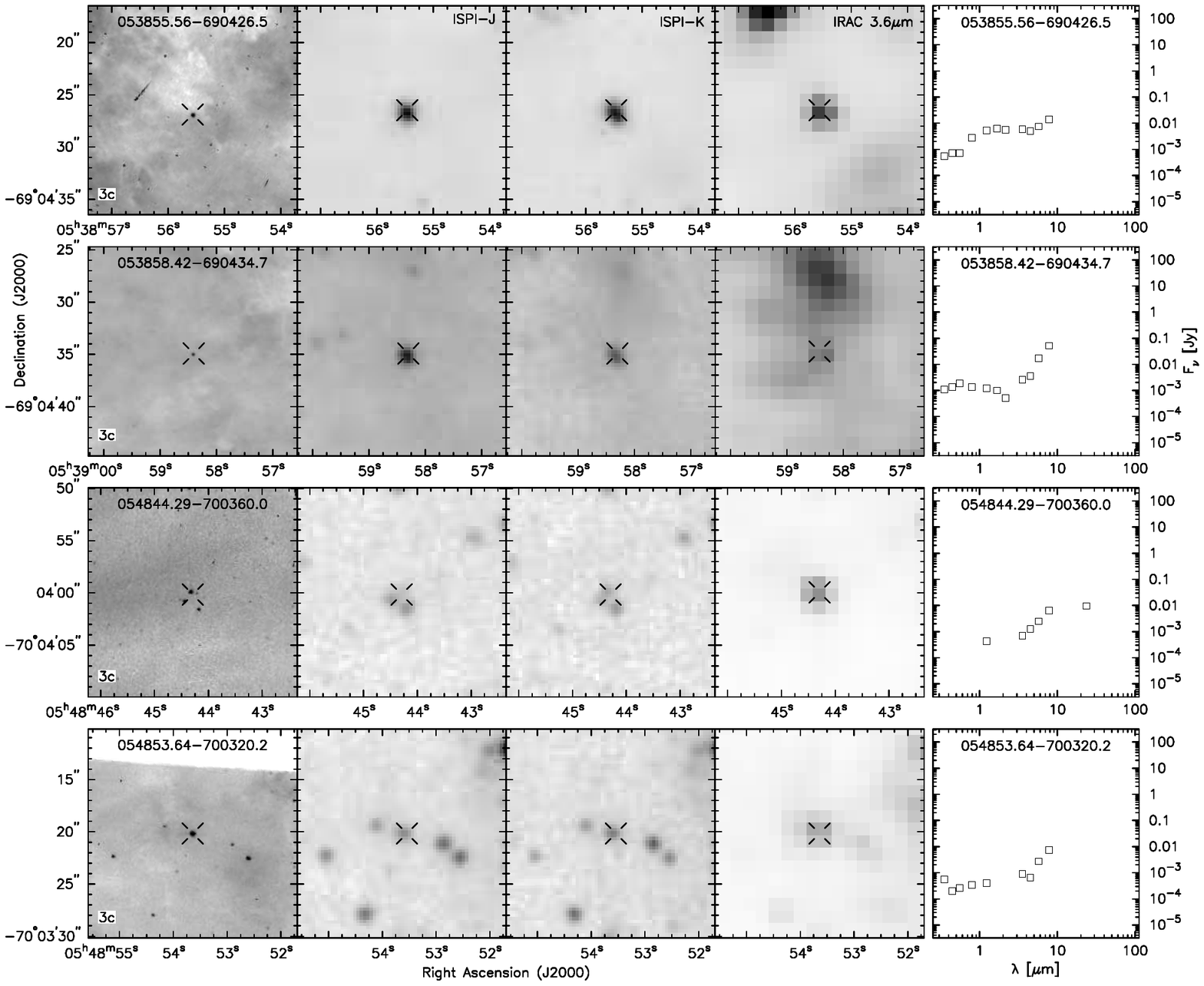}














\clearpage

\begin{figure}
\begin{center}
\includegraphics[scale=0.9]{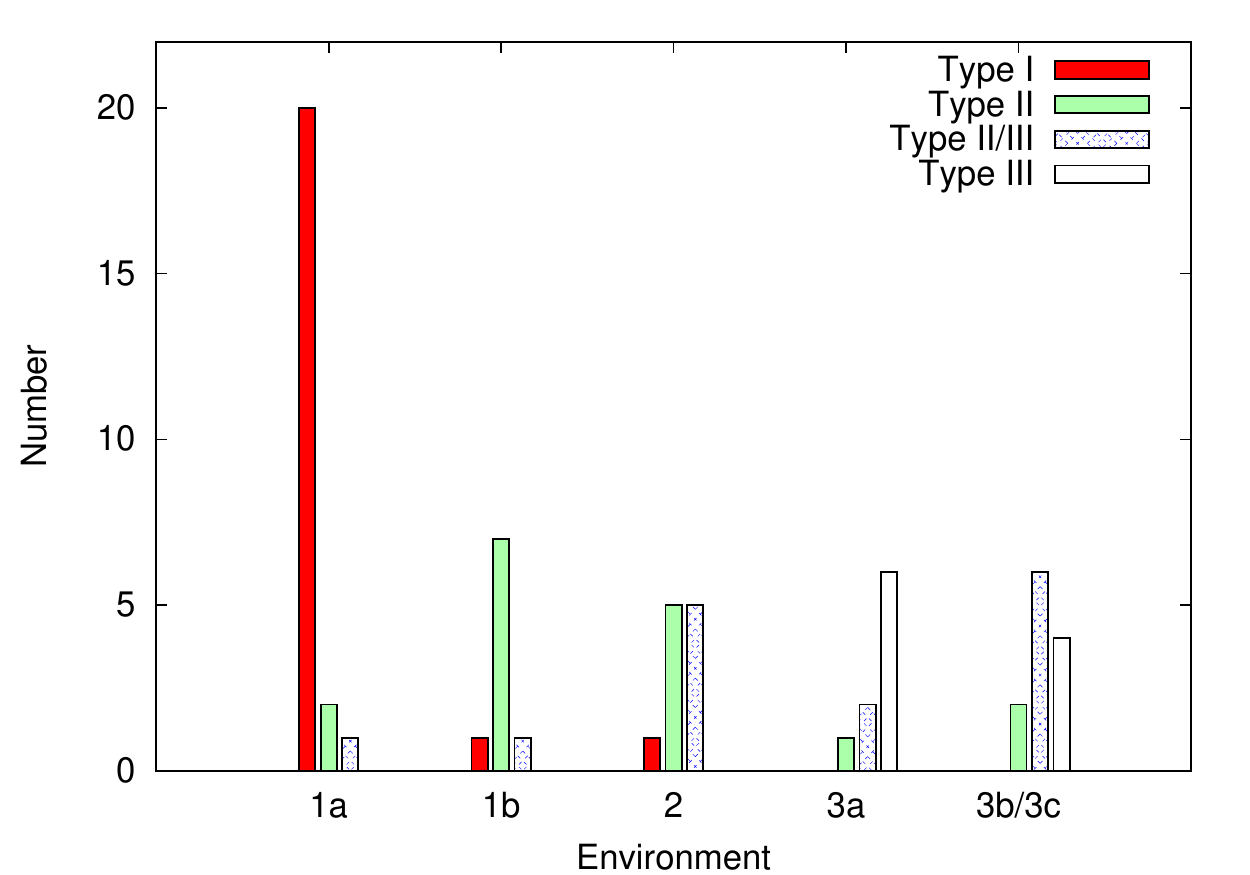}
\caption{A histogram plot showing the correlation of YSO environments and the SED Type classification.
YSOs in dark clouds with no optical counterparts are 1a, YSOs in   
dark clouds with optical counterparts are 1b, YSOs in bright rimmed dust globules are 2, 
YSOs with resolved \ion{H}{2} regions are 3a, YSOs with marginally 
resolved \ion{H}{2} regions are 3b and YSOs with unresolved \ion{H}{2} regions are 3c.}
\end{center}
\end{figure}

\clearpage

\appendix
\begin{deluxetable}{ccl}
\tabletypesize{\scriptsize}
\setlength{\tabcolsep}{0.03in}
\tablecolumns{10}
\tablewidth{0pt}
\tablecaption{List of non-YSOs}
\tablehead{
\colhead{Number} & \colhead{ID} & \colhead{Description}}
\startdata
1 & J045647.11-662459.1 & ~~~~~~Diffuse emission \\
2 & J045658.24-662430.7 & ~~~~~~Diffuse emission \\
3 & J045702.52-662503.3 & ~~~~~~Diffuse emission \\ 
4 & J045726.42-662248.4 & ~~~~~~galaxy \\
5 & J052218.80-675814.6  & ~~~~~~Diffuse emission \\
6 & J053525.90-691428.6 & ~~~~~~galaxy \\
7 & J053550.40-692422.0 & ~~~~~~galaxy \\
8 & J053554.84-691426.6 & ~~~~~~Diffuse emission \\
9 & J053627.62-691434.9 & ~~~~~~Diffuse emission \\
10 & J053708.79-690720.3 & ~~~~~~A bright star \\
11 & J053742.63-690943.6 & ~~~~~~Diffuse emission \\
12 & J053825.21-690405.2 & ~~~~~~Diffuse emission \\
13 & J053844.32-690329.9 & ~~~~~~Diffuse emission with perhaps faint YSOs \\
14 & J053845.99-690930.8 & ~~~~~~Diffuse emission \\
15 & J053848.86-690828.0 & ~~~~~~galaxy \\
16 & J053906.22-692930.9 & ~~~~~~Star on diffuse emission \\
17 & J053949.18-693747.4 & ~~~~~~Two galaxies \\
\enddata
\end{deluxetable}

\end{document}